\documentclass[11pt,letterpaper]{article}
\usepackage{etoolbox}

\usepackage{amsthm}
\usepackage{amsmath,latexsym,amssymb,verbatim,enumerate,graphicx}
\usepackage{hyperref}
\usepackage{color}
\usepackage{enumitem}
\usepackage{booktabs}
\RequirePackage{mathtools}

\newtheorem*{rep@theorem}{\rep@title}
\newcommand{\newreptheorem}[2]{%
\newenvironment{rep#1}[1]{%
 \def\rep@title{#2 \ref{##1} (restatement)}%
 \begin{rep@theorem}}%
{\end{rep@theorem}}}
\makeatother

\newtheorem{thm}{Theorem}
\newtheorem*{thm*}{Theorem}
\newtheorem{definition}[thm]{Definition} 
\newtheorem{prop}[thm]{Proposition}
\newtheorem*{prop*}{Proposition}
\newtheorem{lem}[thm]{Lemma}
\newtheorem*{lem*}{Lemma}

\newtheorem*{fact*}{Fact}
\newtheorem{cor}[thm]{Corollary}
\newtheorem*{cor*}{Corollary}
\newtheorem{con}[thm]{Conjecture}

\newreptheorem{thm}{Theorem}
\newreptheorem{lem}{Lemma}
\newreptheorem{cor}{Corollary}

\def\ba#1\ea{\begin{align}#1\end{align}}
\def\ban#1\ean{\begin{align*}#1\end{align*}}
\newcommand{\subeq}[2]{\begin{subequations} \label{eq:#1} \begin{align}
      #2 \end{align} \end{subequations}}

\newcommand{\ot}{\otimes}
\newcommand{\be}{\begin{equation}}
\newcommand{\ee}{\end{equation}}

\def\QMA{{\sf{QMA}}}

\def\BellQMA{{\sf{BellQMA}}}

\def\SymQMA{{\sf{SymQMA}}}

\def\EXP{{\sf{EXP}}}

\def\BellSymQMA{{\sf{BellSymQMA}}}

\def\NP{{\sf{NP}}}
\def\NEXP{{\sf{NEXP}}}
\def\QMIP{{\sf{QMIP}}}
\def\MIP{{\sf{MIP}}}

\def\MA{{\sf{MA}}}

\def\PSPACE{{\sf{PSPACE}}}

\def\SAT{{\sf{SAT}}}

\def\benum{\begin{enumerate}}
\def\eenum{\end{enumerate}}

\def\nn{\nonumber}

\def\squareforqed{\hbox{\rlap{$\sqcap$}$\sqcup$}}
\def\qed{\ifmmode\squareforqed\else{\unskip\nobreak\hfil
\penalty50\hskip1em\null\nobreak\hfil\squareforqed
\parfillskip=0pt\finalhyphendemerits=0\endgraf}\fi}
\def\endenv{\ifmmode\;\else{\unskip\nobreak\hfil
\penalty50\hskip1em\null\nobreak\hfil\;
\parfillskip=0pt\finalhyphendemerits=0\endgraf}\fi}

\newcommand{\bra}[1]{\langle #1|}
\newcommand{\ket}[1]{|#1\rangle}

\newcommand{\bbR}{\mathbb{R}}
\newcommand{\bbC}{\mathbb{C}}

\newcommand{\id}{\mathbb{I}}

\newcommand{\ben}{\begin{equation}}
\newcommand{\een}{\end{equation}}

\newcommand{\<}{\langle}
\renewcommand{\>}{\rangle}

\def\L{\left}
\def\R{\right}

\def\id{{\operatorname{id}}}
\DeclareMathOperator{\Sep}{Sep}
\DeclareMathOperator{\SepSym}{SepSym}
\DeclareMathOperator{\supp}{supp}

\DeclareMathOperator{\tr}{tr}
\def\be{\begin{equation}}
\def\ee{\end{equation}}
\def\ben{\begin{eqnarray}}
\def\een{\end{eqnarray}}
\def\ot{\otimes}

\def\bei{\begin{itemize}}
\def\eei{\end{itemize}}

\DeclareMathOperator*{\E}{\mathbb{E}}

\def\ep{\varepsilon}
\def\eps{\varepsilon}

\def\cB{{\cal B}}
\def\cD{{\cal D}}
\def\cE{{\cal E}}
\def\cH{{\cal H}}
\def\cM{{\cal M}}

\def\cS{{\cal S}}
\def\cX{{\cal X}}
\def\cY{{\cal Y}}
\def\tA{\tilde{A}}

\mathchardef\ordinarycolon\mathcode`\:
\mathcode`\:=\string"8000
\def\vcentcolon{\mathrel{\mathop\ordinarycolon}}
\begingroup \catcode`\:=\active
  \lowercase{\endgroup
  \let :\vcentcolon
  }

\newcommand{\nc}{\newcommand}
\nc{\rnc}{\renewcommand} \nc{\beq}{\begin{equation}}
\nc{\eeq}{{\end{equation}}} \nc{\bea}{\begin{eqnarray}}
\nc{\eea}{\end{eqnarray}} \nc{\beqa}{\begin{eqnarray}}
\nc{\eeqa}{\end{eqnarray}} \nc{\lbar}[1]{\overline{#1}}
 \nc{\proj}[1]{|#1\rangle\!\langle #1 |} 
\nc{\avg}[1]{\langle#1\rangle}

\nc{\smfrac}[2]{\mbox{$\frac{#1}{#2}$}} \nc{\Tr}{\operatorname{Tr}}
\nc{\ox}{\otimes} \nc{\dg}{\dagger} \nc{\dn}{\downarrow}
\nc{\lmax}{\lambda_{\text{max}}}
\nc{\lmin}{\lambda_{\text{min}}}

\nc{\csupp}{{\operatorname{csupp}}}
\nc{\qsupp}{{\operatorname{qsupp}}} \nc{\var}{\operatorname{var}}
\nc{\rar}{\rightarrow} \nc{\lrar}{\longrightarrow}
\nc{\poly}{\operatorname{poly}}
\nc{\polylog}{\operatorname{polylog}} \nc{\Lip}{\operatorname{Lip}}
\nc{\Om}{\Omega}
\nc{\wt}[1]{\widetilde{#1}}

\def\>{\rangle}
\def\<{\langle}

\def\ve{\varepsilon}

\nc{\glneq}{{\raisebox{0.6ex}{$>$}  \hspace*{-1.8ex} \raisebox{-0.6ex}{$<$}}}
\nc{\gleq}{{\raisebox{0.6ex}{$\geq$}\hspace*{-1.8ex} \raisebox{-0.6ex}{$\leq$}}}

\nc{\vholder}[1]{\rule{0pt}{#1}}
\nc{\wh}[1]{\widehat{#1}}
\nc{\h}[1]{\widehat{#1}}

\nc{\ob}[1]{#1}

\def\beq{\begin {equation}}
\def\eeq{\end {equation}}

\def\be{\begin{equation}}
\def\ee{\end{equation}}

\nc{\eq}[1]{(\ref{eq:#1})} 
\nc{\eqs}[2]{\eq{#1} and \eq{#2}}

\nc{\eqn}[1]{Eq.~(\ref{eqn:#1})}
\nc{\eqns}[2]{Eqs.~(\ref{eqn:#1}) and (\ref{eqn:#2})}

\newcommand{\secref}[1]{Section~\ref{sec:#1}}

\newcommand{\lemref}[1]{Lemma~\ref{lem:#1}}
\newcommand{\thmref}[1]{Theorem~\ref{thm:#1}}
\newcommand{\propref}[1]{Proposition~\ref{prop:#1}}

\newcommand{\corref}[1]{Corollary~\ref{cor:#1}}
\newcommand{\conref}[1]{Conjecture~\ref{con:#1}}

\nc{\region}{\cS\cW}

\def\bz{\mathbf{z}}

\newtoggle{long}
\nc{\longonly}[1]{\iftoggle{long}{#1}{}}
\nc{\shortonly}[1]{\iftoggle{long}{}{#1}}
\nc{\longshort}[2]{\iftoggle{long}{#1}{#2}}
\usepackage{fullpage}
\toggletrue{long}

\begin{document}

\shortonly{
\conferenceinfo{STOC'13,} {June 1–4, 2013, Palo Alto, California, USA.} 
\CopyrightYear{2013} 
\crdata{978-1-4503-2029-0/13/06} 
\clubpenalty=10000 
\widowpenalty = 10000
}

\title{Quantum de Finetti Theorems under Local Measurements with Applications}

\shortonly{\subtitle{Extended Abstract\titlenote{A full version of this paper is available on the arXiv.}}}

\longshort{
\author{Fernando G.S.L. Brand\~ao\footnote{Department of Computer Science, University College London, and National Quantum Information Center of Gdansk. email: {\tt fgslbrandao@gmail.com}}
\and Aram W. Harrow \footnote{
Center for Theoretical Physics, Massachusetts Institute of Technology.  email: {\tt aram@mit.edu}
}}
\date{\today\vspace{-1em}}
}
{\numberofauthors{2} 
\author{
\alignauthor
Fernando~G.S.L.~Brand\~ao\\
	\affaddr{Computer Science \\ University College London, and National Quantum Information Center of Gdansk }\\
\email{fgslbrandao@gmail.com}
\alignauthor
Aram~W.~Harrow\\
\affaddr{Physics \\ MIT}\\
\email{aram@mit.edu}
}}

\maketitle

\begin{abstract}

Quantum de Finetti theorems are a useful tool in the study of correlations in quantum multipartite states. In this paper we prove two new quantum de Finetti theorems, both showing that under tests formed by local measurements in each of the subsystems one can get a much improved error dependence on the dimension of the subsystems. We also obtain similar results for non-signaling probability distributions. We give the following applications of the results to quantum complexity theory, polynomial optimization, and quantum information theory:

\vspace{0.2 cm}

\addtolength{\leftskip}{1.5em} \setlength{\parindent}{-1.5em}

\makebox[1.5em][l]{$\bullet$} We prove the optimality of the Chen-Drucker protocol for 3-$\SAT$, under the \longshort{assumption there is no subexponential-time algorithm for $\SAT$}{exponential-time hypothesis}. In the protocol a prover sends to a verifier $\sqrt{n}\polylog(n)$ unentangled quantum states, each composed of $O(\log(n))$ qubits, as a proof of the satisfiability of a 3-$\SAT$ instance with $n$ variables and $O(n)$ clauses. The quantum verifier checks the validity of the proof by performing local measurements on each of the proofs and classically processing the outcomes. We show that any similar protocol with $O(n^{1/2 - \varepsilon})$ qubits would imply a $\exp (n^{1 - 2\varepsilon}\polylog(n))$-time algorithm for 3-$\SAT$. \par \medskip

\makebox[1.5em][l]{$\bullet$} We show that the maximum winning probability of free games (in which the questions to each prover are chosen independently) can be estimated by linear programming in time $\exp(O(\log|Q| + \log^2|A|/\varepsilon^2) )$, with $|Q|$ and $|A|$ the question and answer alphabet sizes, respectively, matching the performance of a previously known algorithm due to Aaronson, Impagliazzo and Moshkovitz. This result follows from a new monogamy relation for non-locality, showing that $k$-extendible non-signaling distributions give at most a $O(k^{-1/2})$ advantage over classical strategies for free games. We also show that 3-$\SAT$ with $n$ variables can be reduced to obtaining a constant error approximation of the maximum winning probability under entangled strategies of $O(\sqrt{n})$-player one-round non-local games, in which only two players are selected to send $O(\sqrt{n})$-bit messages. \par \medskip

\makebox[1.5em][l]{$\bullet$} We show that the optimization of certain polynomials over the complex hypersphere can be performed in quasipolynomial time in the number of variables $n$ by considering $O(\log(n))$ rounds of the Sum-of-Squares (Parrilo/Lasserre) hierarchy of semidefinite programs. This can be considered an analogue to the hypersphere of a similar known results for the simplex. As an application to entanglement theory, we find a quasipolynomial-time algorithm for deciding multipartite separability. \par \medskip

\makebox[1.5em][l]{$\bullet$} We consider a quantum tomography result due to Aaronson -- showing  that given an unknown $n$-qubit state one can perform tomography that works well for most observables by measuring only $O(n)$ independent and identically distributed (i.i.d.) copies of the state -- and relax the assumption of having i.i.d copies of the state to merely the ability to select subsystems at random from a quantum multipartite state.  \par \medskip

\vspace{0.1 cm}

\addtolength{\leftskip}{- 1.5em}\setlength{\parindent}{0.8 em}

The proofs of the new quantum de Finetti theorems are based on information theory, in particular on the chain rule of mutual information.  The results constitute improvements and generalizations of a recent de Finetti theorem due to Brand\~ao, Christandl and Yard.

\end{abstract}

\shortonly{
}

\section{Background}

A central problem in quantum information theory, quantum computation, and physics in general is to understand \textit{entanglement}: quantum correlations with no counterpart in classical probability theory. An important technique in the study of entanglement are quantum versions of the de Finetti theorem. The latter states that the marginal probability distribution $p^{X_1\ldots X_l}$ on $l$ subsystems of a permutation-symmetric probability distribution $p^{X_1\ldots X_k}$ on $k \geq l$ subsystems is close (within $l(l-1)/k$ in variational distance) to a convex combination of independent and identically distributed (i.i.d.) probability distributions \cite{DF80}. This is a powerful result as it allows us to infer a very particular form for $p^{X_1\ldots X_l}$ merely based on a symmetry assumption on $p^{X_1\ldots X_k}$.  Note we can always make sure this assumption holds true by merely forgetting the order of the $k$ subsystems. Quantum versions of the de Finetti theorem state that a $l$-partite quantum state $\rho^{A_1\ldots A_l}$ that is a reduced state of a permutation-symmetric state on $k \geq l$ subsystems is close (for $k \gg l$) to a convex combination of i.i.d. quantum states, i.e. $\rho^{A_1\ldots A_l} \approx \int \mu(d\sigma) \sigma^{\otimes l}$ for a probability measure $\mu$ on quantum states. 

The quantum version appears very similar to the original de Finetti theorem, but it is much more remarkable. Not only it says that the correlations are arranged in an organized fashion (as a convex combination of i.i.d. states) but also that the state of $l$ subsystems is close to a \textit{separable}, non-entangled, state. A well-known property of entanglement is that it is monogamous: A quantum system cannot be very much entangled with a large number of other systems. The quantum de Finetti theorems provide a quantitative statement for the monogamy of entanglement; in a symmetric state all the subsystems are equally correlated with all the others and so each of them can only be slightly entangled with a few of the others. 

We now know several possible quantum versions of the de Finetti theorem~\longshort{\cite{HM76, Sto69, FLV88, RW89, Wer89, CFS02, KR05, CKMR07, Ren07, NOP09, BCY10b}}{\cite{HM76, KR05, CKMR07, Ren07, NOP09, BCY10b}}. A natural way to quantify the closeness to convex combinations of i.i.d. states is by the trace norm \longonly{\footnote{The trace norm gives the maximum probability of distinguishing two quantum states by arbitrary measurements.}}. In this case \longshort{Christandl, K\"onig, Mitchison, and Renner}{Ref.}~\cite{CKMR07} proved an almost optimal quantum de Finetti theorem: $\rho^{A_1\ldots A_l}$ is $(2d^2l/k)$-close to a convex combination of i.i.d. states in trace norm, with $d$ the dimension of the subsystems, while there are examples where the error is $\Omega(dl/k)$. However in many applications this error is too large to be useful. One possible way forward is therefore to consider other ways of quantifying the approximation rather than the trace norm.

There are two known quantum de Finetti theorems following this idea. The first is the exponential de Finetti theorem of Renner \cite{Ren07}, that achieves an exponentially small error in $k-l$, but only shows that $\rho^{A_1\ldots A_l}$ is close to a convex combination of "almost i.i.d." states, a generalization of i.i.d. states having similar properties with respect to certain statistical tests. The second is the de Finetti theorem proved in Ref.~\cite{BCY10}, which works for $l = 2$ and has an error of $\sqrt{16 \ln(d) /k}$, an exponential improvement on the dimension dependence. The approximation is quantified by the one-way LOCC\longonly{\footnote{The name LOCC stands for local operations and classical communication. See Eq. \eq{LOCC1} for a precise definition of one-way LOCC.}} norm, a variant of the trace norm for bipartite systems in which only measurements implementable by local operations and one-directional classical communication are allowed. Both results have found interesting applications\longshort{: The first to quantum key distribution \cite{Ren05}, quantum hypothesis testing \cite{BP10}, and quantum state tomography \cite{Ren07}; the second to entanglement testing, where it gives a quasipolynomial-time algorithm for determining if a quantum state is entangled or not \cite{BCY10b}, and to quantum complexity theory \cite{BCY10b}. These two results suggest that more quantum versions of the de Finetti theorem might exist. In this paper we show that this is indeed the case. 

It has emerged that some of the properties of entanglement, such as its monogamous character, are shared by more general classes of correlations \cite{MAG06}. A particular interesting example is the class of non-signaling distributions, which are a generalization of the correlations attainable by quantum mechanics. Versions of the de Finetti theorem for non-signaling distributions have also been derived~\cite{CT09, BF09}, although here again the scaling of the error -- linear in the number of possible measurements -- has limited the applicability of the results.
}{.  In particular \cite{BCY10b} gave a quasipolynomial-time algorithm for determining if a quantum state is entangled or not and was used to give the best known bound for the convergence of the Sum-of-Squares hierarchy for approximating the 2-to-$q$ norm of matrices \cite{BHKSZ12}.}

Another way to study quantum entanglement is via its role in operational tasks, e.g. in quantum key distribution and quantum computation. One fascinating case is the role of entanglement in \textit{quantum proof systems}.  \longonly{The goal there is to understand how useful are entangled states for convincing a verifier the truth of a mathematical statement.}  There are many settings, such as interactive or non-interactive protocols, one or multiple provers, and which type of communication is allowed among the provers and the verifier \longonly{(see e.g. \cite{Wat08})}. In this paper we will be concerned with two such settings in particular. The first is $\MIP^*$, in which the provers share entanglement \longonly{(or even general non-signaling correlations)} and are only allowed to communicate with the verifier and not with each other\longonly{~\cite{KM03}}. The second is $\QMA(k)$, meaning non-interactive multiple proof protocols with the assumption that the proofs are \textit{not} entangled\longonly{~\cite{KMY03}}.  \longonly{Here we have the interesting situation where the assumption of not having entanglement among the proofs appears to give extra power to the proof system.  Both settings have been extensively studied in the past (see e.g. \cite{CHTW04, Weh06, NPA08, DLTW08, KRT10, KKMTV11, IKPSY08, IKM09, CGJ09, KV10, IV12} for work on $\MIP^*$/$\QMIP$ and \cite{ABDFS08, BT07, HM10, HM13, CD10, Bra07, BCY10, Bei10, GNN12, CF11, GSU11, McK11, CS11, Per12, SW11} for work on $\QMA(k)$), although there are still many interesting open questions concerning them. }

\section{Results}

The main results of this paper are two new quantum versions of the de Finetti theorem, along with extensions to arbitrary non-signaling distributions. Both are based on a coarser notion of approximation to the target state than the trace norm, but as a pay-off their error scales exponentially better with dimension. The notion of approximation used is that two quantum states are close if they have the same statistics under any local measurements on the subsystems. Our results thus extend the de Finetti bound of Ref.~\cite{BCY10b} to an arbitrary number of subsystems while improving on the error term to depend on the number of measurements instead of the local dimension, generalizing it to general non-signaling distributions, and in some cases providing an explicit rounding scheme.  Among the applications of the new quantum de Finetti theorems we address two problems in quantum complexity theory, each concerning one of the proof systems mentioned above. Below we give a brief description of these applications.

\vspace{0.2 cm}

\noindent \textbf{Multiple Unentangled Proofs:} The first application concerns a protocol due to Chen and Drucker~\cite{CD10} in which a prover sends to a verifier $\sqrt{n}\polylog(n)$ unentangled quantum states, each composed of $O(\log(n))$ qubits, as a proof of the satisfiability of a 3-$\SAT$ instance with $n$ variables and $O(n)$ clauses. The quantum verifier then checks the validity of the proof by performing local quantum measurements on each of the proofs and post-processing the outcomes. This result (building on \cite{ABDFS08}), is surprising since one can convince a verifier the satisfiability of a 3-$\SAT$ instance by sending only $\sqrt{n}\polylog(n)$ qubits! It is a natural question whether the total number of qubits could be decreased even further.  As a direct application of one of the new quantum de Finetti theorems we give strong evidence against any further decrease: We show that any similar protocol with $O(n^{1/2 - \varepsilon})$ qubits, for any $\varepsilon > 0$, would imply a $\exp (n^{1 - 2\varepsilon} \polylog(n))$-time algorithm for 3-$\SAT$. This proves the optimality of the protocol under the \longshort{plausible assumption that there are no subexponential-time algorithms for $\SAT$ \cite{ImpagliazzoPZ98}}{exponential-time hypothesis}. 

A related, but harder, problem is whether $\QMA(2)$ protocols can give at most a quadratic reduction in proof size with respect to $\QMA$ \longonly{\footnote{$\QMA$ is the quantum version of $\NP$. $\QMA(2)$, in turn, is a version of $\QMA$ in which one is given two proofs, with the promise they are not entangled with each other; see section \ref{sec:BellQMA}.}$^{,}$\footnote{By Ref.~\cite{HM13} we know $\QMA(2)$ with constant soundness gives at least a quadratic reduction in proof size relative to $\QMA$, under plausible computational complexity assumptions; see section \ref{sec:BellQMA}.}}. We believe the result we obtain gives evidence that this might be the case and that a suitable quantum version of the de Finetti theorem might be the right tool to show it \longonly{\footnote{See \cite{BCY10b, HM13} for more evidence this might be the case, along with obstacles to prove it.}}.

\vspace{0.2 cm}

\noindent \textbf{Non-local Games:} The second application concerns the computational complexity of non-local games. We give two results in this direction. The first is algorithmic and concerns the class of free games, defined as games in which the questions to each prover are chosen independently. We show that the maximum winning probability of such games can be approximated within additive error $\varepsilon$ by a linear program in time $\exp(O(\log|Q| + \log^2|A|/\varepsilon^2) )$, with $|Q|$ and $|A|$ the question and answer alphabet sizes, respectively. The run-time matches the performance of a different algorithm for the problem due to Aaronson, Impagliazzo and Moshkovitz~\cite{AIM14}\footnote{This algorithm was communicated to us already in 2010, although the result has appeared publicly only in \cite{AIM14}.}. Although this is a purely classical result, we establish it by exploring a connection to non-local games: We show that for any two-player one-round free game, one can find another game on $m$ players such that the maximum winning probability under non-signaling strategies, which can be computed by a linear program \cite{Ito10}, gives a $\sqrt{\frac{\ln|A|}{2m}}$-additive approximation to the maximum winning probability of the original game. Note that since non-signaling strategies are at least as powerful as entangled strategies, the same result holds also for games in which the players share entanglement. 

Using the observation above for entangled strategies, together with a hardness result for free games from \cite{AIM14}, we also show that  3-$\SAT$ on $n$ variables can be reduced to obtaining a \textit{constant error} approximation of the maximum winning probability under entangled strategies of $O(\sqrt{n})$-player one-round non-local games, in which the players communicate $O(\sqrt{n})$ bits all together. Finally, we show how one would be able to establish $\NP$-hardness of approximating the maximum winning probability under entangled strategies of a 4-player one-round game if one could strengthen appropriately one of the new quantum de Finetti theorems of this paper. This gives a new approach to this famous problem, which was only recently resolved~\cite{Vidick13}.

\vspace{0.2 cm}

\noindent \textbf{Polynomial Optimization:} We consider the connection \cite{DohertyPS04, BHKSZ12, DW12} between quantum de Finetti theorems and the optimization over separable states, on one hand, and polynomial optimization and the Sum-of-Squares (Parrilo/Lasserre) hierachy, on the other hand, and prove that the optimization of certain degree-$d$ polynomials over the $n$-dimensional hypersphere can be approximated to error $\varepsilon$ in quasipolynomial-time in the number of variables by considering $O(\log(n)d^3\varepsilon^{-2})$ rounds of the Sum-of-Squares hierarchy of semidefinite programs.
\longonly{This result can be considered as an extension to the hypersphere of similar results for the simplex \cite{PR01}.} Moreover employing the result of \longonly{Chen and Drucker} \cite{CD10}, we show that $\Omega(d^2)$ rounds are necessary to obtain even a constant error-approximation, \longshort{unless there are subexponential-time algorithms for $\SAT$}{assuming the exponential-time hypothesis}.

\vspace{0.2 cm}

\noindent \textbf{Separability Testing:} Another application is to give an algorithm for deciding separability of multipartite states which is quasi-polynomial in the local dimensions of the subsystems. Given a multipartite state $\rho_{A_1, \ldots, A_l}$, we prove one can decide whether it is fully separable or $\varepsilon$-away from separable in time $\exp\left( O \left( \left(\sum_k \ln|A_k|\right)^2  l^2 \varepsilon^{-2}  \right)   \right)$, with distance measured either by the one-way LOCC norm \longonly{\cite{BC11}} or by a multipartite version of the Frobenius norm introduced in \cite{LW12b}. This generalizes the findings of \cite{BCY10b} from bipartite states to general multipartite states\longonly{, and vastly improves on the bound of \cite{BC11}}. 

\vspace{0.2 cm}

\noindent \textbf{Efficient State Tomography:} A final application of the new de Finetti theorems is to quantum state tomography. The starting point is a result due to Aaronson \cite{Aar06}, based on computational learning theory, showing that given an unknown $n$-qubit state one can perform tomography that allows us to compute to good accuracy the statistics of most observables by measuring only $O(n)$ i.i.d. copies of the state. The new de Finetti theorem we prove allow us to relax the assumption of having i.i.d. copies of the state (which can never be fully certified), showing that essentially the same conclusion holds true for arbitrary quantum states, as long as one can selects a few of its subsystems at random and performs the original scheme on them (weakening however the number of subsystems needed from $O(n)$ to $\poly(n)$, of which only $O(n)$ are measured and the rest discarded).

\vspace{0.2 cm}

\noindent \textbf{Notation:} Let ${\cal D}({\cal H})$ be the set of quantum states on ${\cal H}$, i.e. positive semidefinite matrices of unit trace acting on the vector space ${\cal H}$. We say $\rho^{AB} \in {\cal D}(A \otimes B)$ is a $k$-extendible state if there is a state $\tilde \rho^{AB_1\ldots B_k} \in {\cal D}(A \otimes B^{\otimes k})$ such that $\tilde \rho^{AB_j} = \rho^{AB}$ for all $j \in [k]$.  For a multipartite state such as $\rho^{XY}$, we use the convention that omitting subscripts corresponds to taking the partial trace over those systems; e.g. $\rho^X = \tr_Y \rho^{XY}$ in the previous example.  Let $\text{Sep}$($A:B$) denote the set of separable states in ${\cal D}(A \otimes B)$, which is defined to be the convex hull of the states of the form $\rho^A \ot \rho^B$ (product states).  Similarly $\Sep(A^{\ot l})$ is the convex hull of states of the form $\rho_1 \ot \cdots \ot \rho_l$.
We say $\rho^{A_1\ldots A_k} \in {\cal D}(A^{\otimes k})$ is permutation symmetric if $\rho^{A_{\pi(1)}\ldots A_{\pi(k)}} = \rho^{A_1\ldots A_k}$ for any permutation $\pi \in S_k$ (with $S_k$ the symmetric group of order $k$).

A quantum measurement (also called a POVM or positive-operator valued measure) is given by a set of matrices $\{ M_k \}$ such that $M_k \geq 0$ and $\sum_k M_k = I$. We associate to any measurement a map $\Lambda(X)  = \sum_k \tr(M_k X) \ket{k}\bra{k}$, with $\{ \ket{k} \}$ an orthonormal basis. We denote the set of maps associated to measurements by ${\cal M}$. These are also called quantum-classical channels, since they map quantum states to probability distributions.

Let $p(x_1, \dots, x_k | a_{1}, \dots, a_{k}) \in {\cal X}^{\times k} \times {\cal A}^{\times k}$ be a conditional probability distribution. We say it is non-signaling if $p(x_j | a_j)$ is independent of $a_k$ for $k \neq j$. We say $p(x, y | a, b)$ is $k$-extendible if there is a non-signaling distribution \shortonly{\\} $p(x, y_1, \dots, y_k | a, b_{1}, \dots, b_{k})$ which is permutation-symmetric in the $B$ systems\longonly{, i.e. $p(x, \pi^{-1}(y_1), \dots, \pi^{-1}(y_k) | a, \pi^{-1}(b_1), \dots, \pi^{-1}(b_k)) = p(x, y_1, \dots, y_k | a, b_{1}, \dots, b_{k})$ for all permutations $\pi \in S_k$,} and whose marginal is $p(x, y | a, b)$. We call LHV (local hidden variable) the set of conditional probability distributions of the form $p(x, y |a, b) = \sum_l \pi_l q_{l}(x | a) r_{l}(y, | b)$ for a probability distribution $\pi$ and local conditional distributions $q_l, r_l$.


\subsection{Quantum de Finetti Theorems for Local Measurements}

By monogamy of entanglement we expect that a $k$-extendible state $\rho^{AB}$ to be close to a separable state, since the $A$ subsystem is equally correlated to $k$ systems. The next theorem gives a quantitative version of this fact both for entanglement and for non-signaling distributions.

\def\ThmBipartite#1{

\mbox{}

\benum

\item 
Let $\rho^{AB} \in {\cal D}(A \otimes B)$ be a $k$-extendible state and $\mu(m)$ a distribution over quantum operations $\{ \cE_m^{A \rar \tA} \}_{m}$, with $\cE_m^{A \rar \tA} : {\cal D}(A) \rightarrow {\cal D}(\tA)$. Then
\ba
&\min_{\sigma \in \Sep(A:B)}   \max_{\Lambda^B \in {\cal M}} \mathop{\mathbb{E}}_{m\sim \mu}
 \left \Vert \cE_m^{A\rar \tA} \otimes \Lambda^B \left( \rho^{AB} -  \sigma^{AB} \right)  \right \Vert_{1} 
\shortonly{\nn\\} \leq \shortonly{&}
 \sqrt{ \frac{ 2 \ln |\tA| }{k}}.
\label{eq:bipartite-bound-#1}\ea

\item 
Let $\rho^{AB} \in {\cal D}(A \otimes B)$ be a $k$-extendible state, $\mu(m)$ a distribution over operators $\{\cE_m^{A\rar\tA}\}_m$ from $\cD(A)\rar\cD(\tA)$ and $\Lambda^B$ a measurement on $\cD(B)$. Then in time $\poly(|A|, |B|^k)$ a classical computer can compute $\sigma \in \Sep(A:B)$ such that
\be \E_{m\sim \mu}   \left \Vert \cE_m^{A\rar \tA} \otimes \Lambda^B \left( \rho^{AB} -  \sigma^{AB} \right)  \right \Vert_{1}  \leq \sqrt{ \frac{ 2 \ln |\tA| }{k}}.
\ee

\item 
Let $p(x, y | a, b) \in {\cal X} \times {\cal Y} \times {\cal A} \times {\cal B}$ be a $k$-extendible non-signaling conditional probability distribution and let $\mu$ be a distribution over ${\cal A}$. Then
\ba
& \min_{q \in \text{LHV}}   \max_{b \in {\cal B}} \mathop{\mathbb{E}}_{a \sim \mu}   \left \Vert  p(x, y | a, b) - q(x, y | a, b)  \right \Vert_{1}  
\longshort{\leq}{\nn \\ \leq &}
 \sqrt{ \frac{ 2 \ln |X| }{k}}.
\ea
\eenum}

\begin{thm} \label{thm:bipartite}\label{THM:BIPARTITE}
\ThmBipartite{first}
\end{thm}

The de Finetti bound from Ref.~\cite{BCY10b} can be recovered (with an improved constant) as a special case of part 1 of \thmref{bipartite} by choosing the singleton distribution composed of the ideal channel on $A$, since \footnote{The one-way LOCC norm is defined as $\Vert X\Vert_{\text{LOCC}^{\leftarrow}} = \max_{0 \leq M \leq I} \tr(XM)$, with the the maximization over all POVMs $\{M, I - M \}$ that can be realized by local operations and one-way classical communication from $B$ to $A$.}
\begin{equation}
\Vert \rho^{AB} - \sigma^{AB} \Vert_{\text{LOCC}^{\leftarrow}} = \max_{\Lambda\in\cM}
 \Vert (\id \otimes \Lambda)(\rho^{AB} - \sigma^{AB}) \Vert_1.
\label{eq:BCY}\end{equation}
However, \thmref{bipartite} improves on Ref.~\cite{BCY10b} in several ways.  First and most importantly, the error term is independent of the subsystem dimensions of $\rho^{AB}$, and only depends on the output dimension of the family of quantum operations $\{ \cE_m^{A\rar \tA} \}_{m}$, thus yielding nontrivial bounds even if $A$ is infinite-dimensional. Likewise, for non-signaling distributions the bound in part 3 is independent of the number of measurement settings of $p(x, y |a, b)$.   Second, if we think of $k$-extendable states as a relaxation of $\Sep$, then part 2 provides an ``explicit rounding,'' which did not exist in Ref.~\cite{BCY10b}, although we note the caveat that $\sigma$ depends on the measurement $\Lambda^B$.  Third, part 3 generalizes the result to non-signaling distributions.
Note that in part 1, taking system $\tA$ to be classical would yield a special case of part 3, but in the more general case where $\tA$ is a quantum system, parts 1 and 3 are incomparable.  

\longonly{We remark that \eq{bipartite-bound-first} (in the case where $\cE_m=\id$) also follows from \longshort{the work of Yang~\cite{Yang06} using the fact that the entanglement of formation \cite{BDSW96} is upper-bounded by the log of either of the local dimensions, together with a variant of the Pinsker inequality adapted to LOCC$^\leftarrow$~\cite{Piani09}. It also follows from the recent work of Li and Winter \cite{LW12a}}{Refs.~\cite{Yang06,LW12a}}.}

The proof of Theorem~\ref{thm:bipartite} (found in \secref{bipartite-proof}) is more direct and general than the proofs in \cite{BCY10b, Yang06, LW12a}, in particular not making use of entanglement measures in any explicit way.   This enables us to obtain parts 2 and 3 of the theorem\longonly{~(but see the discussion of Conjecture~\ref{con:games} for an example of how the generality of \thmref{bipartite} limits our abilities to further improve it)}.   

We remark that the explicit rounding in part 2 was mostly known only for the variants requiring $k\geq |B|$~\cite{Ren07,CKMR07,NOP09,DW12}, and the previous de Finetti theorems for non-signaling boxes~\cite{CT09, BF09} similarly required $k$ to scale as some power of the local dimension.  This means that the resulting algorithms for approximating $\Sep$ would take time exponential in the dimension.   At first glance it would appear that our \eq{bipartite-bound-first} is incomparable to the previous results; for example, Thm 2 of \cite{NOP09} implies that if $\rho$ is $k$-extendable then there exists $\sigma\in\Sep(A:B)$ with $\|\rho-\sigma\|_1 \leq 2/(1+k/|B|)$.  In \propref{B-dim} (found in \secref{bipartite-proof}), we show how similar bounds can be obtained with our information-theoretic methods.

One other previous work to find $k$ scaling logarithmically with dimension is \cite{BKS12}, which achieves a similar but incomparable bound for measurements with nonnegative matrix elements, together with an efficient rounding scheme.  Ref.~\cite{BKS12} was also an important source of inspiration for the current work\footnote{Update: this result can now be found in Section 3 of \cite{BarakKS14}.}.

The next theorem gives a generalization of the result of \cite{BCY10b} to an arbitrary number of subsystems\longonly{, as well as to non-signaling distributions}.
 
\def\ThmMulti{\mbox{}

\longonly{\benum\item}
  Let $\rho^{A_1\ldots A_k} \in {\cal D}(A^{\otimes k})$ be a permutation-invariant state. Then for every $0 \leq l \leq k$ there is a measure $\nu$ on ${\cal D}(A)$ such that
\shortonly{for every $\Lambda_2, ..., \Lambda_l \in {\cal M}$,}
\ba
&\longonly{\max_{\Lambda_2, \ldots, \Lambda_l \in {\cal M}}}
 \left \Vert \left( \id \otimes \Lambda_2 \otimes \ldots  \otimes \Lambda_l \right) \left( \rho^{A_1\ldots A_l} - \int \nu(d\sigma) \sigma^{\otimes l}  \right)      \right \Vert_1 
\longshort{\leq}{\nn \\ \leq &}
 \sqrt{  \frac{ 2  l^2 \ln |A|}{k - l } }.
\ea
\longonly{\item
Let $p(X_1\cdots X_k | A_1\cdots A_k)$ be a permutation-invariant non-signaling conditional probability distribution (i.e. $p$ is invariant under simultaneous permutation of the $X$ and $A$ systems).  Fix a product distribution $\mu = \mu_1 \otimes \cdots \otimes \mu_k$ on $A_1\times\cdots\times A_k$.  Then for every $0<l<k$ there is a measure $\nu$ on single-system conditional probability distributions such that
\be
\E_{a_1,\ldots,a_l\sim\mu}
\L\| p(X_1\cdots X_l | a_1,\ldots, a_l) - \E_{q\sim \nu} q(X_1|a_1) \ot \cdots \ot q(X_l|a_l) \R\|_1
\leq \sqrt{\frac{2 l^2 \ln |X|}{k-l}}
\ee
\eenum}
}

\begin{thm} \label{thm:multi}
\ThmMulti
\end{thm}

\longshort{
In Ref.~\cite{DF80} Diaconis and Freedman proved that for a permutation-symmetric probability distribution $p_{k}$ on $k$ subsystems, $p_{l}$ is $\frac{l(l-1)}{k}$-close (in variational distance) to a convex combination of i.i.d. probability distributions. Theorem \ref{thm:multi} can be seen as an analogue of this result to quantum states and non-signaling probability distributions. However instead of having a bound which is independent of the dimension, we only have a bound that depends logarithmic on the dimension (and the notion of approximation is weaker than variational distance). It is an interesting question whether this can be improved. Note however that we give in Section \ref{sec:BellQMA} a computational complexity argument that the $k  \geq \Omega(l^2)$ dependency is optimal.

Just as \thmref{bipartite} yielded a stronger version of the BCY result~\cite{BCY10b} as a corollary, \thmref{multi} leads to a multipartite version of \eq{BCY}.  The main difference, apart from considering state on $l$ systems that have symmetric $k$-partite extensions, is that the bipartite LOCC$^{\leftarrow}$ norm is replaced by one in which parties $2,\ldots,l$ measure their systems and communicate the outcomes to party 1, who can then choose a measurement adaptively based on these messages.  This leads to a norm on states that can be thought of as a multipartite generalization of the LOCC$^{\leftarrow}$ norm.  We note that \cite{BC11} also derived a multipartite generalization of \cite{BCY10b} but with an exponentially worse scaling of $k$ with $l$.
}
{In the full version~\cite{BH12a} we also give an analogue of Theorem \ref{thm:multi} for non-signaling distributions. We note this theorem can be considered as an analogue of Diaconis and Freedman's classical de Finetti theorem~\cite{DF80}, however with a logarithmic dependence on the dimension, instead of the alphabet independence achieved in \cite{DF80}.
}

{\bf Proof idea:} The proofs can be found in \secref{bipartite-proof}, but here we sketch the intuition behind part 1 of \thmref{bipartite}.  If $\rho^{AB}$ is $k$-extendible, then we can treat it instead as part of a state $\rho^{AB_1\cdots B_k}$ with $\rho^{AB} = \rho^{AB_i}$ for each $i$.  First examine systems $A$ and $B_1$.  If $\rho^{AB_1}$ is approximately product, then $\rho^{AB}$ is approximately separable and we are done.  If not, then the correlations between $A$ and $B_1$ mean that conditioning on $B_1$ will reduce the entropy of $A$.  Then we can examine the mutual information between $A$ and $B_2$ conditioned on $B_1$.  Again, if this is small, then we have a nearly separable state and can stop, and if not, then we can condition on $B_2$ and further reduce the entropy of $A$.  Since the initial entropy of $A$ is at most $\log |A|$, this process is effective as long as $k\gg \log|A|$, which is a benefit of our information-theoretic approach over most previous versions of the de Finetti theorem.  The main difficulty is that conditioning on a system does not work (indeed is not defined) if that system is quantum.  This introduces the main subtlety, which is that we need to measure all of the $B_i$ systems, but then use the outputs of the measurement to reason about the properties of the state before it was measured.  Since the post-measured state is automatically separable, this requires some care.

{\bf Relation to previous work:} Our approach here is in a sense a throwback to the original quantum de Finetti theorem of Caves, Fuchs and Schack~\cite{CFS02}.  They too applied an informationally complete measurement to the quantum systems and applied a classical de Finetti theorem to the resulting outcomes.    However, they could only prove that the approximation tends to zero as $k\rar \infty$ with no control on the rate, while our error bounds are optimal or nearly-optimal in many settings.  There are a few differences in our approaches.  
Ref.~\cite{CFS02} does full tomography on the quantum state, while we only consider a specific measurement $\Lambda^B$, which is what enables us to have error scaling with the log of the dimension, but also which limits us only bounding error in the 1-LOCC norm.
More importantly, Ref.~\cite{CFS02} treats the classical de Finetti theorem as a black box, which means that approximation errors from de Finetti can become uncontrolled when inverting state tomography.  Our approach essentially reproves the de Finetti theorem (as we discuss further in Section 4 of our followup paper~\cite{BH-product}) and by keeping track of the states along the way, we avoid having to carry out such a state-reconstruction task.  In a way, this was also the idea behind \cite{KR05}, which gave the first finite quantum de Finetti theorem.  Our approach differs from \cite{CFS02,KR05} and most of the other previous work by avoiding state reconstruction and informationally complete measurements altogether; instead our ``rounding'' step is achieved by replacing a state $\rho^{AB}$ with $\rho^A \ot \rho^B$ in a situation where this is guaranteed to cause little error.  

\subsection{Non-Local Games: Algorithms and Hardness Results}\label{sec:games}

One application of Theorem \ref{thm:bipartite} is to the computational complexity of non-local games. A multiprover game is played between a set of cooperative players/provers, who are not allowed to communicate with each other, and a referee/verifier who interrogates the provers to decide if they win the game.  \longshort{ In a one-round game, for example, the verifier chooses questions to each prover at random and checks the answers obtained from the provers in order to decide whether to accept or not. Even though the provers cannot communicate with each other, they can agree on a common strategy in order to win the game with the maximum probability possible.

Multiprover games have had a central role in computational complexity theory. In a seminal paper Babai, Fortnow, and Lund proved $\NEXP = \MIP$ \cite{BFL91}, with $\MIP$ the class of languages having multi-prover interactive proof with a polynomial number of provers, rounds, and bits exchanged between the provers and the verifier in each round. Building on \cite{BFL91}, it was then proven in \cite{ALMSS98, AS98} that it is $\NP$-hard to approximate to constant error the maximum winning probability of a two-player one-round game (with the input size given by the total number of questions to the players and their answers). This hardness result is equivalent to the celebrated PCP theorem \cite{ALMSS98, AS98}, which has a pivotal role in hardness of approximation results (see e.g. \cite{AB09}). }
{Multiprover games have had a central role in computational complexity theory.}

It is natural to allow the players to share correlations that might assist them in winning the game with a higher probability.  \longshort{While it is easy to see that shared randomness is of no help, it}{It} has been known since the seminal work of Bell\longonly{~\cite{Bell64}} that entanglement might help the players to win with a probability strictly larger than with a purely classical strategy. One can even consider stronger correlations than the ones allowed by quantum mechanics, such as arbitrary non-signaling correlations. Games in which the players can use entanglement (or more general non-signaling correlations) are known as non-local games\longonly{, since the extra shared resources allow the players to sometimes use strategies that cannot be reproduced by local ones (i.e. strategies only using shared randomness and local actions)}.  Upper bounds on the maximum winning probability of a one-round game under classical strategies are known as Bell inequalities, and non-local strategies that beat these bounds are known as Bell inequality violations.  Such violations of Bell inequalities are central in the foundations of quantum mechanics as they can be implemented experimentally to show that nature cannot be described by a local hidden variable theory\longonly{~\cite{ADR82}}. 

\longshort{Given the usefulness of multiprover games to computational complexity theory and of non-local games to the foundations of quantum mechanics, it is interesting to study how difficult it is to compute}{We consider here the complexity of computing}
the \textit{entangled} value of the game, defined as the maximum probability of winning the game using entanglement, or the \textit{non-signaling} value of the game, defined as the optimal probability under non-signaling strategies.   By contrast, the maximum winning probability under classical strategies is called the {\em classical} value of the game.

\longshort{Although {\em a priori} computing the entangled value of a game requires optimizing over a large set, in some cases this can be easier.  Indeed, for unique games, the best known algorithms for the classical value~\cite{AroraBS10} run in time $\exp(n^\eps)$ (with $0<\eps<1$ depending on the desired degree of approximation), whereas the entangled value of the game can be estimated in polynomial time using semidefinite programming~\cite{KRT10} (or exactly calculated for the special case of XOR games~\cite{CHTW04}). These two classes of games could be taken as evidence that the estimation of the entangled value is generally easier than of the classical value.  However if one is interested in a high-accuracy estimation this turns out not to be true. Kempe, Kobayashi, Matsumoto, Toner, and Vidick proved that it is $\NP$-hard to approximate to an inverse polynomial (in the size of the game) the entangled value of one-round 3-prover games \cite{KKMTV11} (see also \cite{IKPSY08, IKM09, CGJ09}). Recently in a beautiful development Ito and Vidick \cite{IV12} proved that it is $\NP$-hard, under quasi-polynomial reductions (improved to a polynomial-time reduction in Ref.~\cite{Vidick13}), to approximate the entangled value of 3-prover games with polynomially many rounds even to constant error. The result \cite{IV12} has a more elegant formulation in terms of interactive proof systems: It shows that $\NEXP \subseteq \MIP^*$, with $\MIP^*$ the analogue of $\MIP$ in which the provers share entanglement \cite{KM03}. The maximum probability of non-signaling strategies, in turn, can always be computed efficiently by linear programming~\cite{Ito10}.


Probably the biggest open question in this area is to determine the computational complexity of approximating the entangled value of one-round games to constant accuracy (although recent work of Vidick~\cite{Vidick13} has now resolved this in all but the case of two players). There are two reasons why this is a particular interesting setting. The first is the fact that the PCP theorem can be stated as the $\NP$-hardness of approximating the classical value of one-round games to constant accuracy. Thus an analogous result for the entangled value could be interpreted as a version of the PCP theorem in the presence of entanglement. Second, in Bell inequality violation experiments, which are one-round non-local games, one can only obtain a constant-accuracy approximation to the true violation due to experimental error. Therefore it is important to understand how efficiently one can estimate to constant error the maximum violation of a Bell inequality, since this the most experimentally relevant approximation scale. One of our goals here is to propose a new approach to address this problem.}
{Concerning entangled games, it was proven in \cite{KKMTV11} that it is $\NP$-hard to approximate to an inverse polynomial (in the size of the game) the entangled value of one-round 3-prover games. Recently in a beautiful development,
 Vidick~\cite{Vidick13} (building on \cite{IV12}) proved that it is $\NP$-hard to approximate the entangled value of one-round 3-prover games even to constant error. The maximum probability of non-signaling strategies, in turn, can always be computed efficiently by linear programming, and $\MIP^{\text{ns}}(2,1)=\PSPACE$, where $\MIP^{\text{ns}}$ denotes two-prover one-round interactive proof with no-signaling provers~\cite{Ito10}.}

A particular class of games that we will consider are the so-called \textit{free games}, defined as games in which the questions to each of the players are chosen independently from the questions to the other players~\cite{CCL90}. A famous example from physics is the CHSH game. The fact that the verifier cannot coordinate questions suggests that the computation of the maximum winning probability of such games should not be as hard as for general games. And indeed Bellare, Feige and Killian proved that the analogue of $\MIP$ for poly-round free games is equal to $\PSPACE$~\cite{BFK95}, while Aaronson, Impagliazzo and Moshkovitz~\cite{AIM14} proved that the classical value of one-round free games with questions to the two provers in $Q\times Q$ and answers in $A_1\times A_2$ can be \longonly{simulated to within error $\ve$ by AM (Arthur-Merlin) proofs with an $O(\log |Q|+\log(|A_1|\cdot|A_2|)/\ve)$-bit message from Arthur to Merlin and an $O(\log|A_1|\log|A_2|/\ve)$-bit message from Merlin to Arthur. As a result, the value of such games can be} estimated in time $\poly(\log |Q|)\exp(\log|A_1| \log|A_2| / \ve)$. They also gave a matching hardness of approximation result for free games, showing that one can reduce 3-$\SAT$ on $n$ binary variables to computing $\omega_{c}(G)$ to within constant additive error for 2-player one-round free games with $\exp(O(\sqrt{n}))$-sized answer alphabet\longonly{\footnote{There are suggestive similarities between this result and results about $\QMA(2)$ and variants thereof; see \secref{BellQMA} and \cite{HM13}.}}.  

As a corollary of Theorem \ref{thm:bipartite} we will prove that the classical value of free games can be computed efficiently by linear programming, matching the run-time of the algorithm of \cite{AIM14}. Moreover, we will also derive a non-trivial hardness of approximation result for the \textit{entangled} value of free games by importing to the case of entangled strategies the hardness of approximation result for the classical value of free games from \cite{AIM14} . Finally we will show how a conjectured strengthening of Theorem \ref{thm:bipartite} would yield an alternate proof of the $\NP$-hardness of obtaining a constant error approximation of $\omega_e$ for four-player one-round games.


Before we turn to the precise statement of the main result of this section let us give a \longshort{more formal}{brief} definition of non-local games\longshort{.
\begin{definition}}{:}
We define a $m$-prover game $G(m, \pi, V)$ by two parameters $\pi$ and $V$:
\longshort{\benum \item}{(i)}
 $\pi$ is a probability distribution on $Q_1 \times \ldots  \times Q_m$ for finite sets $Q_1,\ldots,Q_m$\longshort{.\item}{; and (ii)} $V$ is a predicate on $Q_1 \times \ldots  \times Q_m \times A_{1} \times \ldots  \times A_{m}$ for finite sets $A_1,\ldots, A_m$.
\longonly{\eenum}
The sets $Q_i$ and $A_i$ consist of the possible questions and answers, respectively, for player $i$. The predicate $0 \leq V(a_1, \ldots, a_m | q_1, \ldots, q_m) \leq 1$ is the pay-off function of the answer $(a_1, \ldots, a_m)$ given the question $(q_1, \ldots, q_m)$. 
\longonly{\end{definition}}

The \textit{classical} value of the game $G$ is given by
\ba
&\omega_{c}(G(m, \pi, V)) := \max_{a_1, \ldots, a_m}
\shortonly{\\ & \nn}
 \sum_{q_1, \ldots, q_m} \pi(q_1, \ldots, q_m) V(a_1(q_1), \ldots,  a_m(q_m)| q_1, \ldots, q_m),
\ea
where the maximum is over all functions $a_j : Q_j \rightarrow A_j$.

The \textit{entangled} value of the game, in turn, is given by
\begin{eqnarray}
&&\omega_{e}(G(m, \pi, V)) 
\longshort{\nn \\ &:=&}{:=} \sup \sum_{q_1, \ldots, q_m} \pi(q_1, \ldots, q_m) 
\sum_{a_1, \ldots, a_m}\shortonly{\\ \nn  &&}
V(a_1, \ldots, a_m | q_1, \ldots, q_m)  \bra{\psi} M_{a_1 | q_1}^1 \otimes \ldots . \otimes M_{a_m | q_m}^m \ket{\psi},
\end{eqnarray}
where the supremum is over states $\ket{\psi}$ of arbitrary dimension and arbitrary POVMs 
\longshort{\begin{equation}}{$}
\{ M_{a_1 | q_1}^1 \}_{a_1\in A_1}, \ldots, \{ M_{a_m | q_m}^m \}_{a_m\in A_m},
\longshort{\end{equation}}{$}
with $\sum_{a_k \in A_k} M_{a_k | q_k}^k = I$ for every $q_k \in Q_k$ and $k \in [m]$.

Finally, the \textit{non-signaling} value of the game $G$ is defined as
\begin{eqnarray}
&&\omega_{ns}(G(m, \pi, V))  \longshort{\nn \\ &:=&}{:=}
  \max \sum_{q_1, \ldots, q_m} \pi(q_1, \ldots, q_m)
\shortonly{\\ \nn &&} \sum_{a_1, \ldots, a_m} V(a_1, \ldots, a_m | q_1, \ldots, q_m)   p(a_1, \ldots, a_m | q_1, \ldots, q_m),
\end{eqnarray}
where the maximum is over all non-signaling probability distributions $p(a_1, \ldots, a_m | q_1, \ldots, q_m)$.



\def\CorGame#1{
\mbox{}
\benum 
\item Let $G(2, \pi, V)$ be a two-player one-round non-local free game with $\pi$ a product probability distribution on $R \times Q$ and $V$ a predicate on $R \times Q \times A \times B$. Then there is a $(m+1)$-player one-round non-local game $\overline G(m+1, \overline \pi, \overline V)$ with $\overline \pi$ a probability distribution on $R \times Q_1 \times \ldots  \times Q_m$, with $|Q_k| = |Q|$ for $k \in [m]$, and $\overline V$ a predicate on $R \times Q_1 \times \ldots  \times Q_m \times A \times B_1 \times \ldots  \times B_m$, with $|B_k| = |B|$ for $k \in [m]$, such that
\ba \label{relationomegaens#1}
\omega_{c}(G)  = \omega_c(\overline G)
&\leq \omega_{e}(\overline G) \leq \omega_{ns}(\overline G) 
\shortonly{\nn\\&}\leq \omega_{c}(G) +  \sqrt{\frac{\ln|A|}{2m}}.
\ea

\item For a free game $G(2, \pi, V)$ there is a linear-programming relaxation of size $|R||A|\left(|Q||B|\right)^{\frac{\ln|A|}{2 \varepsilon^2}}$ for computing $\omega_{c}(G)$ to within additive error $\varepsilon$.

\item One can reduce 3-$\SAT$ on $n$ variables to computing $\omega_{e}(G)$ to within constant additive error for $O(\sqrt{n})$-player one-round non-local games with answer alphabet size of $\exp(O(\sqrt{n}))$ in which only two players are asked questions.

\eenum 
}

\begin{cor} \label{cor:hardnessbell} 
\CorGame{}
\end{cor}

See \longshort{\secref{proof-games}}{the full version~\cite{BH12a}} for the proof.

\longonly{We note that it is trivial to prove either a version of part 3 of Corollary \ref{cor:hardnessbell} in which the answer alphabet size is $2^n$ (in which case even one prover is clearly enough), or one in which the answer alphabet size is constant but one has $n$ provers, or one with $\sqrt{n}$ provers and alphabet size $2^{\sqrt{n}}$ in which {\em all} provers respond.  However, in our result, the total number of bits sent is $O(\sqrt{n})$.}


Part 2 of Corollary \ref{cor:hardnessbell} follows directly from part 1 and the fact that $\omega_{ns}$ can be computed by linear-programming. \longonly{This gives a new algorithm matching the performance of the algorithm due to Aaronson, Impagliazzo and Moshkovitz~\cite{AIM14}.} Part 3 of Corollary \ref{cor:hardnessbell} follows from part 1 and the hardness of approximation result of Ref.~\cite{AIM14} for free games.\longonly{ 

} Part 1 in turn gives a generic relation between the classical value of a free game, on one hand, and the quantum and non-signaling values of a modified game with more players, one the other hand. The idea of adding more players is to try to immunize the original game from entanglement (or general non-signaling correlations) by adding extra consistency tests that forces the entanglement between the players to have a specific form. Indeed the new game with $m+1$ players consists of playing the original game with player one and one of the remaining $m$ players chosen at random. This essentially allows us to consider a two-player game where the provers can only share an $m$-extendible state (or $m$-extendible non-signaling conditional distribution). Then by Theorem \ref{thm:bipartite} we obtain that this $m$-extendible state cannot be much better than a separable state or a local hidden variable distribution (which themselves are no better than just having shared randomness).\longonly{ The crucial aspect of Theorem \ref{thm:bipartite} used here is that the error term only depends on the number of \textit{outcomes} (which is given by the number of possible answers of the non-local game in question), and not on the dimension of the entangled state or on the number of different POVMs in the family in the quantum case (or the number of measurement settings in the non-signaling case). The idea of immunizing entanglement by introducing more players is not new and was used before by Kempe \textit{et al} \cite{KKMTV11} to prove the hardness of estimating the entangled value within error inverse polynomial in the size of the game. 

More generally, it was observed by Terhal, Doherty, and Schwab \cite{TDS03} that $m$-extendible states cannot violate any Bell inequality with fewer than $m$ measurements for Bob (and an arbitrary number of measurements for Alice). In contrast Theorem \ref{thm:bipartite} shows that a non-signaling $m$-extendible conditional distribution can violate a Bell inequality associated to a free game (an example of which is the CHSH inequality) with an arbitrary number of measurements, each with $M$ possible outcomes, by at most $\frac{1}{2} \sqrt{\frac{2\ln(M)}{m}}$. This is an instance of the concept of monogamy of entanglement (which is known to hold true for non-signaling distributions as well~\cite{CT09}), in this case to the non-locality of quantum states (i.e. the maximum possible violation of a Bell inequality). Note that to be $\varepsilon$-close to a separable state in trace norm (thus having similar statistics under general quantum measurements) one must consider $m$-extendible states with $m = \Omega(|B|/\varepsilon)$, with $|B|$ the dimension of the $B$ subsystem \cite{CKMR07}. The monogamy of non-locality we find here, in comparison, has a bound that is \textit{independent} of the dimension of the state.}

Finally let us mention a conjecture whose validity would imply the $\NP$-hardness of estimating $\omega_{e}$ to within constant error for 4-player one-round games. The conjecture is the following strengthening of Theorem \ref{thm:bipartite}.

\begin{con}\label{con:games}
Let $\rho^{AB} \in {\cal D}(A \otimes B)$ be a $k$-extendible state and $\mu(m)$ a distribution over quantum operations $\{ \cE_m^{A\rar \tA} \}_{m}$, with $\cE_m^{A\rar\tA} : {\cal D}(A) \rightarrow \cD(\tA)$. Then
\longshort{\be}{$$}
\min_{\sigma \in \Sep(A:B)}  \mathop{\mathbb{E}}_{m \sim \mu} \max_{\Lambda^B \in {\cal M}}    \left \Vert \cE_m^{A\rar \tA} \otimes \Lambda^B \left( \longshort{\rho^{AB} - \sigma^{AB}}{\rho-\sigma} \right)  \right \Vert_{1}
\leq \sqrt{ \frac{ 2 \ln |\tA| }{k}}.
\longshort{\label{eq:conj-bound}\ee}{$$}
\end{con}

The difference with Theorem \ref{thm:bipartite} is that the order of the expectation over $\mu$ and the maximization over measurements $\Lambda^B$ is reversed.  \longshort{It is easy to check that one would be able to carry through the proof of part 1 of Corollary \ref{cor:hardnessbell} given in \secref{proof-games} for general games (of course only for the relation of $\omega_e$ and $\omega_c$). The fact that we would be able to prove $\NP$-hardness for $4$-player games would then follows from the combination of this stronger version of Eq. (\ref{relationomegaens}) with a recent version of the PCP theorem due to Khot and Safra, in the language of two-prover one-round games \cite{KS11}. 

We have written \eq{conj-bound} in a way that is meant to parallel \eq{bipartite-bound-first} from \thmref{bipartite}, with a consequence that systems $A$ and $B$ are treated very differently.  However, 
the conjecture could equivalently be restated in a more symmetric form.  If we explicitly include the maximization over $\mu$, then the LHS \eq{conj-bound} becomes $\sup_\mu \min_\sigma [\E_{m\sim \mu} \max_{\Lambda^B} \| \cdots \|_1]$.  Observe that the term inside the $[\cdots ]$ is linear in $\mu$ and convex in $\sigma$; indeed, it is a seminorm of $\sigma$. Thus, we can use Sion's minimax theorem~\cite{Sion} and reverse the order of the $\sup_\mu$ and $\min_\sigma$.  At this point the $\sup_\mu \E_{m\sim \mu}$ become superfluous, and we can replace the pair with simply a maximum over maps $\cE^{A\rar \tA}$.   Thus, \conref{games} could equivalently be stated as
\be \min_{\sigma\in\Sep(A:B)} \max_{\cE^{A\rar \tA}} \max_{\Lambda^B\in \cM}
\left\| \cE^{A\rar \tA} \ot \Lambda^B(\rho^{AB} - \sigma^{AB})\right\|_1
\leq \sqrt{\frac{2 \ln |\tA|}{k}}.\ee

Although the conjecture is consistent with all the examples of states we are aware of, we note that a proof would have to follow a very different approach to the one used in Theorem~\ref{thm:bipartite}, as it cannot apply to non-signaling distributions.  The reason is that the quantum version of the conjecture would imply that $\NEXP\subset\MIP^*(4,1)$, and the no-signaling version would imply that $\NEXP \subset \MIP^{\text{ns}}(4,1)$, but this latter class is contained in $\EXP$\footnote{The proof that $\MIP^{\text{ns}}(\poly,\poly)\subseteq \EXP$ is an easy application of linear programming (essentially the no-signaling constraints are linear constraints on an exponential-sized prover strategy) which appears not to have been published anywhere.   Ref.~\cite{KKMTV11} attribute it to a personal communication from Daniel Preda, and Ref.~\cite{Ito10} builds on this approach to show that $\MIP^{\text{ns}}(2,1)\subseteq\PSPACE$ by finding a way to parallelize the LP in the 2-prove 1-round case.}.  A scaled down version of this argument shows that the no-signaling version of \conref{games} would imply that $\sf{P} = \NP$.  It is an interesting open question to find a more direct counter-argument, such as an example of a $k$-extendable no-signaling distribution whose difference from LHV distributions can be detected by correlated measurements.

Thus, despite the superficial similarity of (the quantum version of) \conref{games} with our \thmref{bipartite}, any proof will need to find features of quantum states that are not shared by no-signaling distributions.  In this respect the hypothesis testing approach of Refs. \cite{BCY10b, LW12a} might be a  promising route. }
{We remark that the non-signaling analogue of Conjecture~\ref{con:games} is false, and since \thmref{bipartite} applies equally well to non-signaling distributions, any proof of the conjecture will require new techniques.}

\subsection{Optimality of Chen and Drucker's\shortonly{\\} Multiple-Proof Protocol for 3-$\SAT$} \label{sec:BellQMA}

One first application of \thmref{multi} is to unentangled multiple proof systems. 

Given a 3-$\SAT$ formula with $n$ variables and $O(n)$ clauses, what is the minimum proof that can convince a verifier the formula is satisfiable? Under the exponential time hypothesis~\cite{IP01} \longonly{-- which says 3-$\SAT$ cannot be solved in subexponential time --} $\Omega(n)$ bits are required, i.e. it is believed one cannot do anything substantially better than just write down the $n$-bit satisfying assignment. What if we can send a quantum state as a proof to a verifier who has a quantum computer to check its validity?
\longshort{Perhaps we could pack more information into the quantum state so that}{Could} $o(n)$ qubits \longonly{would} be enough to convince the verifier?  It turns out that assuming a quantum version of the exponential time hypothesis -- namely that to solve 3-$\SAT$ takes exponential time even on a quantum computer\longonly{ (see e.g. \cite{BBBV97} for the oracle version of this claim)} -- $\Omega(n)$ qubits are required~\cite{MW05}. 

\begin{sloppypar}
Quantum mechanics allows us to add a new twist to this question. What if we want to convince a quantum verifier by sending a quantum state to her, but with the promise that parts of the quantum state are not entangled with each other? In this case the argument of Ref.~\cite{MW05} does not apply anymore and at least we do not have any implausible consequence for having a sublinear proof. And indeed Aaronson, Beigi, Drucker, Fefferman, and Shor \cite{ABDFS08}\longonly{ (building on \cite{BT07})} proved that $\sqrt{n}\polylog(n)$ unentangled quantum states, each of $\log(n)$ qubits, are enough to convince a quantum verifier that a 3-$\SAT$ instance with $n$ variables and $O(n)$ clauses is satisfiable. 

The result of \cite{ABDFS08}  was strengthened in two directions: First Harrow and Montanaro \cite{HM10} proved that \textit{two} unentangled proofs, each of $\sqrt{n}\polylog(n)$ qubits, 
are sufficient. Second Chen and Drucker \cite{CD10} showed that $\sqrt{n}\polylog(n)$ identical unentangled quantum proofs of $O(\log(n))$ qubits each are sufficient to convince even a verifier who measures each of the proofs separately and postprocesses the outcomes in order to decide whether to accept or not.
\end{sloppypar}

To state the main result of this section we define a few quantum complexity classes\longonly{ (see Section \ref{sec:proof-bellqma} for formal definitions)}. The first is a natural quantum analogue of $\NP$ (more precisely of $\MA$). Let $\QMA_n(c, s)$ be the class of problems such that: (i) for "yes" instances there is a quantum proof composed of $n$ qubits that makes the verifier, who has access to polynomial quantum computation, to accept with probability at least $c$; and (ii) for "no" instances every proof is accepted with probability at most $c$. Let $\QMA_{n}(m, c, s)$ be the analogue of $\QMA$ in which instead of one quantum proof the verifier receives $m$ quantum proofs, each of $n$ qubits, with the promise that they are not entangled with each other\longonly{~ \cite{KMY03}}. 

\begin{sloppypar}
Further let $\BellQMA_{n}(m, c, s)$ be an analogue of $\QMA_{n}(m, c, s)$ in which the verification procedure is restricted to applying independent measurements to each of the $m$ proofs and then post-processing the outcomes classically \cite{ABDFS08}. The name of the class comes from the fact that the verifier is basically constrained to apply a Bell test as his verification procedure. Finally let $\BellSymQMA_{n}(m, c, s)$ be the analogue of $\BellQMA_{n}(m, c, s)$ in which all the $m$ proofs are promised to be identical (possibly mixed) states; this corresponds to the set $\SepSym$ from \cite{HM13}.   See \secref{proof-bellqma} for formal definitions of these classes.
\end{sloppypar}

With this notation the Chen-Drucker result can be stated as showing the containment of 3-$\SAT$ with $n$ variables and $O(n)$ clauses in $\BellSymQMA_{\log(n)}(\sqrt{n}\polylog(n), 1-2^{-\Omega(\sqrt{n})}, 1/\poly(n))$ \cite{CD10}. (An analogous, and incomparable, result holds for $\BellQMA$ also follows from \cite{CD10}.)
A corollary of Theorem \ref{thm:multi} is that this is essentially optimal, i.e. the square-root improvement found for the total proof size is all there is if we restrict ourselves to $\BellSymQMA$ protocols.   It is an open question whether an analogous optimal result holds for $\BellQMA$.

\def\CorBellQMA{
\mbox{}
\benum 
\item $\BellSymQMA_{n}(m, c, s) \subseteq \QMA_{10n^2m^2/\varepsilon^2}(c, s + \varepsilon)$.
\item For every $\varepsilon > 0$ and $c- s = \Omega(1)$, there is no $\BellSymQMA_{O(\log(n))}(n^{\frac{1}{2} - \varepsilon}, c, s)$ protocol for 3-$\SAT$ with $n$ variables and $O(n)$ clauses, unless 3-$\SAT$ can be solved in $\exp(n^{1 - 2 \varepsilon} \polylog(n))$ time.
\item $\BellQMA_{n}(m, c, s) \subseteq \QMA_{10n^2m^4/\varepsilon^2}(c, s + \varepsilon)$.
\item 
$\QMA_{\poly(n)}(\frac 23, \frac 13) = \BellQMA_{\poly(n)}(\poly(n), \frac 23,\frac 13)$
\eenum
}
\begin{cor} \label{cor:bellqma} 
\CorBellQMA
\end{cor}
See \longshort{\secref{proof-bellqma}}{the full version~\cite{BH12a}} for the proof.

In \longshort{\cite{Bra07, BCY10b}}{\cite{BCY10b}} it was shown that $\BellQMA(m)$ is contained in $\QMA$ for a constant number of provers $m$. \corref{bellqma} strengthens the containment to even to a polynomial number of provers.  \longonly{This gives a new characterization of the class $\QMA$ and shows that the only advantage (in the regime where $c-s\geq 1/\poly(n)$) that $\BellQMA$ protocols can offer is a polynomial reduction in the proof size, such as in the protocol of \cite{CD10}.}

{\em Remark:} In fact we can prove something slightly stronger than \corref{bellqma}.  Instead of Bell measurements, where $k$ parties individually measure their systems and send the results to a referee, we can handle a slightly larger class of measurements.  Our proofs apply equally to the setting where $k-1$ parties measure their systems and send classical messages to the last party, who can choose a measurement adaptively based on these messages.  This will follow from the fact that in part 1 of \thmref{multi}, we can leave one subsystem unmeasured.  To keep the exposition simple, we will not formally state this improved version of \corref{bellqma}.

\subsection{Polynomial Optimization and \longshort{Sum-of-Squares}{SOS} Proofs} \label{sec:sos}

Another application of our main theorems is to classical algorithms for maximizing polynomials over $\bbC^n$. The concepts of $k$-extendable and separable states turn out to correspond naturally to SDP hierarchies for polynomial optimization, and thus we are able to prove convergence of these hierarchies for polynomials that correspond to LOCC measurements.  This connection was first established by Doherty, Parrilo and Spedalieri~\cite{DohertyPS04}, and was more recently made quantitative for general polynomials over the unit sphere in $\bbR^n$ by Doherty and Wehner~\cite{DW09, DW12}.  

In this section, we consider the problem of maximizing real-valued polynomial functions over the complex unit sphere $S^{2n-1}\subset\bbC^n$.  More precisely, we consider polynomials of $z_1,\ldots,z_n,\bar z_1,\ldots,\bar z_n$ that are bihomogenous of degree $d,d$ (i.e. homogenous of degree $d$ in the $z_1,\ldots,z_n$ and homogenous of degree $d$ in the $\bar z_1,\ldots, \bar z_n$).  This problem is closely related\longonly{~\cite{DP09}} to optimization over the real unit sphere, though not always identical~\cite{CobosKP00}.   When $d>1$, this is generally NP-hard; see \cite{Klerk08}.  A promising general-purpose approximation scheme is to use an SDP hierarchy invented independently by Parrilo~\cite{Parrilo00} and Lasserre~\cite{Lasserre01}; see also \cite{DZ13} for a recent review of the complexity-theoretic properties of this hierarchy.  To define the hierarchy, we introduce some notation.   Let $\bbC[\bz,\bar\bz] := \bbC[z_1,\ldots,z_n,\bar z_1,\ldots, \bar z_n]$ denote complex polynomials in $n$ variables, let $\bbC[\bz, \bar\bz]_{d,d}$ denote the set of bihomogenous polynomials of degree $d,d$, and let $\bbC[\bz, \bar\bz]_d^*$ denote the set of Hermitian linear functionals from $\bbC[\bz, \bar\bz]_d$ to $\bbR$.  Here we will consider only {\em Hermitian} linear functionals $L$, meaning that
$L[\prod_{j=1}^n z_j^{a_j}\bar z_j^{b_j}] =  L[\prod_{j=1}^n z_j^{b_j}\bar z_j^{a_j}]$ for any $a_1,\ldots,a_n,b_1,\ldots,b_n$.

If $p(\bz) \in \bbC[\bz,\bar \bz]_{d,d}$ and $k\geq d$, then we can upper bound $\max_{z\in S^{2n-1}} p(z)$ with the following SDP:
\begin{subequations}\label{eq:SoS-SDP}\ba
\max L(p) & \text{ such that} \\
& L\in\bbC[\bz, \bar \bz]_{k,k}^* \label{eq:SoS-moment} \\
& L(1)=1 \label{eq:SoS-norm}\\
& L(q\bar q)\geq 0 
\longshort{&}{\quad} \forall q \in \bbC[\bz,\bar \bz]_{k,0} \label{eq:SoS-pos}\\
& L((z_1\bar z_1 + \ldots+z_n\bar z_n)q)=L(q) 
\longshort{&}{\nn\\&\qquad}\forall q\in\bbC[\bz,\bar \bz]_{k-1,k-1} \label{eq:SoS-unit}
\ea\end{subequations}

Here \eq{SoS-norm} and \eq{SoS-pos} are constraints that any collection of moments should satisfy (with \eq{SoS-moment} enforcing linearity), while \eq{SoS-unit} expresses the $\sum_{i=1}^n |z_i|^2=1$ constraint (and can in general be replaced with any polynomial constraint; see \cite{Parrilo00,Lasserre01,DZ13}). 
\longonly{To see that \eq{SoS-SDP} is an SDP, observe that \eq{SoS-unit} is a linear constraint and \eq{SoS-pos} is equivalent to the constraint that the moment matrix $M(L)\geq 0$, where the entries of $M(L)$ are indexed by monomials in $\bbC[\bz,\bar \bz]_{k}$ and are defined by $M(L)_{\alpha,\beta} := L(\bar z^\alpha z^\beta)$.}  We can interpret this SDP as replacing the maximum over $S^{2n-1}$ by a maximum over probability distributions over $S^{2n-1}$  (which of course changes nothing), and in turn approximating this by considering only the moments of order $\leq k$.  The dual of \eq{SoS-SDP} is
\subeq{SoS-dual}{ \min \lambda & \text{ such that} \\
& \lambda - p = \L(\sum_{i=1}^n z_i\bar z_i - 1\R)q_0 + \sum_{i=1}^m q_i\bar q_i \\
& q_0\in\bbC[\bz,\bar \bz]_{k-1,k-1} \\
&  q_1,\ldots,q_m \in\bbC[\bz,\bar \bz]_{k,0}}
which can again be seen to be an SDP.  This can be thought of as ``proving'' that $p(x)\leq \lambda$ by using the fact that $p(x)-\lambda$ is a sum of squares of polynomials; hence this SDP is also called the ``sum-of-squares'' hierarchy.

Under reasonable assumptions, as $k$ grows this SDP converges to $\max_{z\in S^{2n-1}} p(z)$ as $k$ grows~\cite{Parrilo00,Lasserre01}.  However, since the effort to compute \eq{SoS-SDP} or \eq{SoS-dual} grows exponentially with $k$, it is important to determine the rate at which this convergence takes place.   This rate is generally well-understood for optimizations over the simplex, but less is known for the sphere~\cite{Klerk08}.

At first glance, the sum-of-squares hierarchy may appear unrelated to the quantum de Finetti theorems studied in this paper.  However, the space $\bbC[\bz]_{k}$ is isomorphic to the symmetric subspace of $(\bbC^n)^{\ot k}$.  Moreover, the relaxation in \eq{SoS-SDP} is tight in the cases when $L$ approximates the evaluation functional (i.e. $L(p) = p(z)$ for some $z\in \bbC^n$) on degree-$d$ polynomials which is analogous to the $d$-body marginals being approximately product.  Indeed, this connection has been explored in \cite{DohertyPS04}, where the sum-of-squares hierarchy was used to prove that $k$-extendable states are approximately separable for sufficiently large $k$, and in \cite{BHKSZ12}, where this connection was used to find cases in which the sum-of-squares hierarchy yielded a good approximation of the $2\rar 4$ norm of a matrix.

To make the connection more explicit, we define, for any convex set $K$, the support function of $K$ by
\longshort{\be}{$} h_K(x) := \sup_{y\in K} \langle x, y\rangle.
\longshort{\ee}{$}
For matrices $x,y$ we define $\langle x,y\rangle := \tr x^\dag y$.   
Let $M$ be a one-way LOCC operator of the form
\be  M = \sum_{i_2,\ldots,i_l} P_{i_2,\ldots, i_l} \ot Q_{2,i_2} \ot \cdots \ot Q_{l,i_l},
\label{eq:one-way-M}\ee
with $0\leq P_{i_2,\ldots, i_l}\leq I$ for each $i_2,\ldots,i_l$ and $0 \leq \sum_{i_j} Q_{j,i_j}\leq I$ for each $2\leq j\leq l$.  Define $\Sep(A^{\ot l})$ to be the convex hull of $\rho_1 \ot \cdots \ot \rho_l$ over $\rho_1,\ldots,\rho_l\in \cD(A)$.
A variant of \thmref{multi} implies that the SOS hierarchy can give a good approximation of
\be h_{\Sep(A^{\ot l})}(M)
\label{eq:multiquadratic-q}\ee
The quantity in \eq{multiquadratic-q} can be thought of as the solution to polynomial optimization problem, or more precisely, a multiquadratic optimization problem.  Given $\ket{z^{(1)}},\ldots,\ket{z^{(l)}}\in \bbC^n$,  define
\be p(z^{(1)},\ldots,z^{(l)}) := \bra{z^{(1)}} \ot \cdots \ot \bra{z^{(l)}} M 
 \ket{z^{(1)}} \ot \cdots \ot \ket{z^{(l)}}. \label{eq:multiquadratic-p}\ee
Then \eq{multiquadratic-q} is equal to the maximum of \eq{multiquadratic-p} over all collections of unit vectors $z^{(1)},\ldots,z^{(l)}$.  

There is an alternate interpretation in terms of polynomial optimization, albeit of some not-very-natural polynomials.  Define $z \in \bbC^{nl}$ to be the vector obtained by concatenating $z^{(1)},\ldots,z^{(l)}$, so that $p(z)$ is a degree-$l,l$ polynomial in $z$.  Maximizing $p(z)$ over unit vectors $z$ can be seen (using a convexity argument) to yield \eq{multiquadratic-q} times the normalization factor $l^{-l}$.

As a result, we immediately obtain a bound on the ability of the sum-of-squares hierarchy to approximate certain polynomials over the complex hypersphere.  

\begin{cor} \label{cor:sos}
Let $p$ be a multiquadratic polynomial of the form in \eq{multiquadratic-p} with $M$ described by \eq{one-way-M}.  Then
\begin{equation} \label{eq:maxoverr}
\max_{\Vert z^{(1)} \Vert_2 = \cdots = \Vert z^{(l)} \Vert_2= 1} p(z^{(1)},\ldots,z^{(l)})
\end{equation}
can be computed to within additive error $\varepsilon$ by $O( \log(n) l^4/\varepsilon^2)$ levels of the sum-of-squares hierarchy, in time $\exp(O(l^4 \log^2(n)/\ep^2))$
\end{cor}

The proof is in \secref{proof-separability}.

Note that the result of Chen and Drucker \cite{CD10} implies that $\log(n) l^{2 - o(1)}$ levels of the sum-of-squares hierarchy are not sufficient to compute even a constant-error approximation to (\ref{eq:maxoverr}), for general $p$ of the form described in the corollary, unless there is a subexponential time algorithm for 3-$\SAT$.   It is an open question whether \corref{sos} can be improved to replace the $l^4$ with $l^2$ to match this bound.  (Recall from \secref{BellQMA} that the Chen-Drucker result cannot be substantially improved without contradicting the exponential-time hypothesis.)

\longonly{There is also more direct evidence that \corref{sos} cannot be improved to yield a PTAS for polynomial optimization over the unit sphere.  Ref.~\cite{BRSW11} proved that for any $n$, there exists a local measurement $M$ (derived from a Bell inequality) on $n\times n$ systems such that 
\be 
\frac{\tr(M\Phi_n)}{h_{\Sep}(M)} \geq \Omega\L(\frac{n}{\log^2(n)}\R),\ee
with $\Phi_n$ the projector onto the $n$-dimensional maximally entangled state. Since $\rho := \frac{1}{k} \Phi_n + (1-\frac{1}{k})\frac{I}{n}$ is $k$-extendable, it follows that the $k$-extendable approximation can make multiplicative errors as large as $\Omega(\frac{n}{k\log^2(n)})$.  Intriguingly, the example of \cite{BRSW11} is based on the unique games problem.   This suggests that using de Finetti theorems to give algorithms for unique games, as suggested by \cite{BHKSZ12}, will need to take advantage of the PPT condition in addition to merely the $k$-extendability property.  The only previous evidence that using the PPT condition gives an asymptotic improvement over mere $k$-extendability was given by \cite{NOP09}.
}

\subsection{Testing Multipartite Separability} \label{sec:Separability}

Another application of part 1 of  \thmref{multi}, closely related to section \ref{sec:sos}, is to the quantum separability problem, a well-studied problem in quantum information theory of both theoretical and practical interest\longonly{~\cite{Ioa07}}. Given a multipartite state $\rho^{A_1 \cdots A_l}$ we say it is fully separable if 
\longshort{\be}{$}
\rho^{A_1 \cdots A_l} = \sum_{j} p_j \sigma_j^{A_1} \otimes \ldots \otimes \sigma_j^{A_l},
\longshort{\ee}{$}
for a probability distribution $\{ p_j \}$ and quantum states $\sigma_j^{A_i}$. 

The goal in the weak-membership problem for separability is to decide whether a given multipartite state $\rho^{A_1\cdots A_l}$ is separable or if it is $\eps$-away from any separable state, given the promise that one of the two alternatives holds true. In fact one has a family of problems depending on which norm we choose to quantify the distance of quantum states. We consider two choices of norms. The first is the one-way LOCC norm, defined as
\longshort{\be}{$}
\Vert X \Vert_{\text{LOCC}^{\leftarrow}} := \max_{\Lambda_2, \ldots, \Lambda_l} \Vert \id \otimes \Lambda_2 \otimes \ldots \otimes \Lambda_l (X) \Vert_1.
\longshort{\ee}{$}
The name comes from the interpretation of norm as $\max_{M} \tr(MX)$, with $M$ any POVM element that can be implemented by parties $2,\ldots,l$ measuring their systems locally and communicating the outcome to party 1, who then performs a measurement dependent on the information received. Therefore we have one-directional communication from all the parties to party 1.

The second is a multipartite version of the Forbenius norm recently introduced by Lancien and Winter~\cite{LW12b}:
\longshort{\be}{$}
\Vert X \Vert_{2(l)} := \sqrt{ \sum_{I \subseteq [l]} \tr \left| \tr_{I} X  \right|^2 }.
\longshort{\ee}{$}

\def\CorSeparability#1{
\begin{sloppypar}
For some $c>0$, the Sum-of-Squares hierarchy solves the weak membership problem for separability for the norm
$\Vert * \Vert_{\text{LOCC}^{\leftarrow}}$ in time
\longshort{\be}{$}
\exp\left(c\left( \sum_j\log |A_j| \right)^2  l^2 \varepsilon^{-2} \right).
\longshort{\label{eq:sep-runtime-#1}\ee}{$}
\longshort{In turn, the Sum-of-Squares hierarchy solves the weak membership problem for separability for the norm $\Vert * \Vert_{2(l)}$ in time 
\be}{ and time $}
\exp\left(c  \left( \sum_j\log  |A_j| \right)^2  (18)^{l/2}l^2 \varepsilon^{-2}
\right).
\longshort{\ee}{$ for $\Vert * \Vert_{2(l)}$.}
\end{sloppypar}}
\begin{cor} \label{cor:separability} 
\CorSeparability{first}
\end{cor}
\longonly{See \secref{proof-separability} for the proof.}

We note this gives a generalization of the result of \cite{BCY10b}, which proved the same result for bipartite quantum states. \longonly{A early generalization of \cite{BCY10b} to multipartite states was given in \cite{BC11}; however there only a bound of 
\be
\exp\left(c \log |A_1| \cdots \log |A_l| l^{2l - 1} \varepsilon^{-2(l - 1)} \right) 
\ee
was obtained for the running time of the algorithm. }

\subsection{Pretty-Good Tomography in Permutation-Symmetric States}

A final application of part 1 of Theorem \ref{thm:multi} is to quantum state tomography, in which one obtains a description of an unknown quantum system by making measurements on the system. In quantum state tomography one tries to obtain a classical description of an unknown quantum state in the form of a density matrix for the state. By performing sufficiently many measurements of a sufficiently large number of different measurement settings one can obtain an arbitrarily good approximation of the true quantum state. Typically one considers a situation in which one has access to many i.i.d. copies of an unknown quantum state, and one performs measurements on those copies in order to learn the identity of the quantum state. Mathematically we can model this situation as saying that the global quantum state is of the form 
\longshort{\be \label{convexofiid}}{$}
\omega_{n} = \int \sigma^{\otimes n} \mu(d \sigma),
\longshort{\ee}{$}
for an unknown measure $\mu$ on quantum states. \longonly{ However the assumption of having many i.i.d. copies of an unknown state cannot always be ensured, and in many situations it simply does not hold true. It is thus an important task to try to relax this requirement. It has long been realized \cite{CFS02} that quantum de Finetti theorems are exactly the right tool here.
Instead of having to assume that $\omega_{n}$ has the form given by Eq. (\ref{convexofiid}), one can merely assume that $\omega_n$ is the reduced state of a larger permutation-symmetric state $\omega_{n+k}$. Then for $k$ sufficiently large $\omega_n$ will be close to a convex combination of i.i.d. states. The point is that one can easily ensure the latter situation by selecting $n$ subsystems at random from the $n+k$ available ones. }{Our work will allow this i.i.d. assumption to be relaxed.  Indeed this was one of the original motivations for quantum de Finetti theorems~\cite{CFS02}.}

The state of affairs is more complicated once complexity is taken into account.  Since a quantum state of $l$ qubits has $4^l$ parameters, reconstructing it generally requires $2^{O(l)}$ different measurement settings. However in many cases most of these parameters do not correspond to relevant questions. For instance, in order to predict expectation values of single-qubit observables, then a linear number of parameters suffices. \longonly{Is there a way to explore this intuition in order to construct more efficient tomographic schemes? }

One beautiful result \longshort{in this direction}{that exploits this intuition} was obtained by Aaronson in Ref.~\cite{Aar06}, using tools from computational learning theory\longonly{~\cite{KV94}}, and can be roughly stated as follows: Given an arbitrary distribution ${\cal M}$ over measurements and an unknown quantum state on $l$ qubits, $O(l)$ measurements settings are sufficient to get a density matrix which, with high probability over the measurement choice from ${\cal M}$, agrees with the expectation of the true quantum state up to small error. \longonly{Thus a linear -- in the number of qubits of the state -- number of measurement settings are enough to get a density matrix which gives a good estimate to the statistics of the true state for almost all choices of measurements; one can perform a "pretty-good" tomography just with a linear number of measurement settings.   The formal statement of Aaronson's result is as follows, restated slightly in order to facilitate our later extension of the result.

\begin{lem}[Theorem 1.3 of \cite{Aar06}] \label{lem:learn-aaronson}
Let $\omega_{m+n}\in\cD(\cH^{\ot m+n})$ be a state of the form
$$\omega_{m+n} = \int \nu(\text{d}\rho) \rho^{\ot m+n},$$
for a probability measure $\nu$ on $\cD(\cH)$.
Let $\cM$ be a distribution over two-outcome measurements
on $\cH$ and $\cE = (E_1,\ldots,E_m)$ a training set of
independently sampled measurements from $\cM$.  Suppose we measure the
first $m$ systems of $\omega$ according to $\cE$ and obtain outcomes
$B=(b_1,\ldots,b_m)\in\{0,1\}^m$.
For any outcome $B$, we will choose a hypothesis state 
\be \sigma_B := \arg\min_\sigma
\sum_{i=1}^m (\tr (E_i\sigma)-b_i)^2.\ee
   Then there exists a constant $K>0$ such that if
\be m \geq \frac{K}{\gamma^4\ve^2}
\L(\frac{\log|\cH|}{\gamma^4\ve^2}\log^2\frac{1}{\gamma\ve}
+ \log\frac{1}{\delta}\R),\ee
then with probability at least $1 - \delta$ the post-measured state $\tilde \omega_n$ satisfies
\begin{equation}
 \tilde \omega_n = \int \rho^{\otimes n} \mu(d\rho),
\end{equation}
where the measure $\mu$ only has non-zero support on states $\rho$ such that
\begin{equation}
\mathop{\text{Pr}}_{E \in {\cal M}} \left[ \left | \tr(E \rho) - \tr(E \sigma_B)   \right | > \gamma   \right] \leq \varepsilon.
\end{equation}
\end{lem}
}

A limitation of Aaronson's result \cite{Aar06}, common of other tomographic schemes as well, is the assumption that one is given several i.i.d. copies of the unknown quantum state. Here too one could try to apply the standard quantum de Finetti theorems \cite{KR05, CKMR07, Ren07} to find a way around this assumption. However since the error in those depend polynomially on the dimension of the state, one would obtain a non-trivial result only if one would select subsystems at random from a state of $2^{O(l)}$ subsystems, which is not a reasonable assumption. Theorem \ref{thm:multi} allows us to circumvent this problem.

\shortonly{In the full version, we state and prove a generalization of Aaronson's result~\cite{Aar06} in which i.i.d. input states are replaced with states that are merely permutation symmetric.}
\longonly{\begin{cor} \label{cor:learning} \mbox{}
Let $\omega_{m + n+k} \in {\cal D}({\cal H}^{\otimes m + n+k})$ be a permutation-symmetric state, let ${\cal M}$ be a distribution over two-outcome measurements on ${\cal H}$, and let ${\cal E} = (E_1, \ldots, E_m)$ be a training set consisting of $m$ measurements drawn independently from ${\cal M}$.  Suppose we discard the last $k$ systems, measure the
first $m$ systems of $\omega$ according to $\cE$ and obtain outcomes
$B=(b_1,\ldots,b_m)\in\{0,1\}^m$.
For any outcome $B$, we will choose a hypothesis state 
\be  \label{sigmaB}
\sigma_B := \arg\min_\sigma
\sum_{i=1}^m (\tr (E_i\sigma)-b_i)^2.\ee
Fix error parameters $\varepsilon, \eta, \gamma, \nu > 0$.   Suppose that (for some universal constant $K>0$) we have
\ba 
m &\geq \frac{K}{\gamma^4\ve^2}
\L(\frac{\log|\cH|}{\gamma^4\ve^2}\log^2\frac{1}{\gamma\ve}
+ \log\frac{1}{\delta}\R),\\
k &\geq \frac{4 (m + n)^2 \ln |{\cal H}|}{\nu^2}.
\ea
Then with probability at least $1 - \delta$ the post-measured state $\tilde \omega_n$ satisfies
\begin{equation} \label{approximationlearning}
\max_{\Lambda_1, \ldots, \Lambda_n} \left \Vert  \Lambda_1 \otimes \ldots  \otimes \Lambda_n \left(  \tilde \omega_n - \int \rho^{\otimes n} \mu(d\rho) \right)     \right \Vert_1 \leq \nu,
\end{equation}
with the maximum over quantum-classical channels $\Lambda_1, \ldots, \Lambda_n$. Here the measure $\mu$ only has non-zero support on states $\rho$ such that
\begin{equation}
\mathop{\text{Pr}}_{E \in {\cal M}} \left[ \left | \tr(E \rho) - \tr(E \sigma_B)   \right | > \gamma   \right] \leq \varepsilon
\end{equation}
\end{cor}

\longonly{The proof of \corref{learning} follows immediately from part 1 of \thmref{multi} and \lemref{learn-aaronson}.}

Let us say a few words about the interpretation of the result. Suppose we had Eq.~(\ref{approximationlearning}) with $\nu = 0$. Then
\be
\tilde \omega_n = \int \rho^{\otimes n} \mu(d\rho),
\ee
with $\mu$ a measure with non-zero support only on states $\rho$ that, for \textit{most} measurements on ${\cal M}$, gives approximately the same statistics as any state $\sigma_B$ compatible with the observed data (in the sense that it satisfies Eq. (\ref{sigmaB})). Therefore any state $\sigma_B$ compatible with the measured data can be used correctly to infer the statistics of future measurements, with high probability over the choice of the observable. 
For non-zero $\nu$ we have a similar situation. While the state $\tilde \omega_n$ might be very far away from a convex combination of i.i.d. in trace norm, if we only consider the statistics of local measurements on the $n$ subsystems, then, up to error $\nu$, we have the same conclusions as in  the case of $\nu = 0$. 

\longonly{The price we have to pay for being able to relax the assumption of having i.i.d. copies of the state is that instead of starting from $O(\log |{\cal H}|) + n$ copies of the state, now we need a global state with $O\left((n+\log|{\cal H}|)^2 \log |{\cal H}| \right)$ subsystems (of which we only measure $O(\log |{\cal H}|)$ of them). The main point is that this is still polynomial in the number of qubits of the unknown state one wants to learn. 

We note that while this approach gives an efficient alternative for tomography of states on a large number of qubits in what concerns the number of measurements needed, it says nothing about the computational complexity of finding the hypothesis state $\sigma_B$. As noted in \cite{Aar06}, it is an interesting problem to determine for which classes of states one can obtain $\rho_v$ efficiently.
}}

\section{\longshort{Proof of \thmref{bipartite} and \propref{B-dim}}{Proofs}}
\label{sec:bipartite-proof}

We will prove \thmref{bipartite} by information-theoretic
techniques, inspired by \cite{BKS12} and Lemma 4.5 of \cite{RT12}.  We will also state and prove \propref{B-dim}, establishing an alternate bound that depends only on the dimension of the $B$ system.

\longonly{Given two quantum states $\rho, \sigma \in {\cal D}({\cal
  H})$, we define the quantum relative entropy (or quantum
Kullback-Leibler divergence) as  
\begin{equation}
S(\rho || \sigma) := \tr(\rho(\ln(\rho) - \ln(\sigma))).
\end{equation}

Given a bipartite state $\rho^{AB} \in {\cal D}(A \otimes B)$ we define the mutual information as 
\begin{equation}
I(A:B)_{\rho} := S(\rho^{AB} || \rho^A \otimes \rho^B).
\end{equation}

Given a tripartite qqc state of the form $\rho^{ABK} := \sum_k p_k \rho^{AB}_k \otimes \ket{k}\bra{k}^K$ we define the conditional mutual information as
\begin{equation}
I(A:B|X)_{\rho} := \sum_k p_k I(A:B)_{\rho_k}.
\end{equation}

The mutual information satisfies the following properties that will be useful in the proof:

\begin{lem} \mbox{}
\benum 
\item \text{Chain Rule:} 
\be I(A:BX) = I(A:X) + I(A:B|X)
\label{eq:chain}\ee
\item \text{Monotonicity under Local Operations:} Let $\pi_{AB} = \id \otimes \Lambda(\rho^{AB})$, then
\begin{equation}
I(A:B)_{\pi} \leq I(A:B)_{\rho}
\end{equation}
\item \text{Pinsker's Inequality:} 
\begin{equation}
I(A:B)_{\rho} \geq \frac{1}{2} \Vert \rho^{AB} - \rho^A \otimes \rho^B \Vert_1^2.
\label{eq:pinsker}\end{equation}
\eenum
\end{lem}

(The absence of the usual $\ln(2)$ factor in \eq{pinsker} is because
of our convention that entropies are measured in ``nats,'' i.e. with
logs taken base $e$.) 

We are now ready to prove \thmref{bipartite}:

\begin{repthm}{thm:bipartite}
\ThmBipartite{second}
\end{repthm}
}

\begin{proof}
\longshort{The three parts of the theorem have similar proofs.  

\noindent \textit{Part 1:}}{In this extended abstract, we describe only the proof of part 1.}

Define the states
\ba
\pi_{\tA B_1\ldots B_kM} &:=  \E_{m\sim \mu} \pi_m \otimes \ket{m}\bra{m}^M
\label{eq:pi-def} \\
\pi_m^{\tA B_1 \ldots B_k} & := 
\left( \cE_m^{A\rar \tA} \otimes
  \Lambda_1^{B_1} \otimes \ldots  \otimes \Lambda_k^{B_k}
\right)(\rho^{AB_1\ldots B_k}),
\nn \ea
with $\cE_m^{A\rar \tA}$ quantum operations from $A$ to $\tA$,
$\Lambda_i$ quantum-classical channels, and $\ket{m}$ a
classical label for which quantum operation $\cE_m^{A\rar \tA}$ was
applied. Repeatedly applying the chain rule\longonly{ \eq{chain}}, we find
\longshort{\begin{equation}}{$}
I(\tA:B_1\ldots B_k |M) = I(\tA:B_1|M) + I(\tA:B_2|M B_1) + \ldots  + I(\tA:B_k | M B_1 \ldots  B_{k-1}).
\longshort{\end{equation}}{$}
Now we maximize over measurements and obtain
\longshort{\begin{align} \label{eq:expansionCMI}
&  \max_{ \Lambda_{1} ,\ldots, \Lambda_{k}  \in {\cal M} } I(\tA:B_1\ldots B_k |M)_{\pi}= \\
 & \max_{   \Lambda_{1} ,\ldots, \Lambda_{{k-1}}  \in {\cal M} }
 \left( I(\tA:B_1|M)_{\pi} + \ldots  + 
I(\tA:B_{k-1} | M B_1 \ldots  B_{k-2})_{\pi} + \max_{ \Lambda_{k} \in {\cal M} }  I(A:B_k | M B_1 \ldots  B_{k-1})_{\pi}  \right). \nonumber 
\end{align}}
{\begin{multline} \label{eq:expansionCMI}
 \max_{ \Lambda_{1} ,\ldots, \Lambda_{k}  \in {\cal M} } I(A:B_1\ldots B_k |M)_{\pi}= \\
\max_{   \Lambda_{1} ,\ldots, \Lambda_{{k-1}}  \in {\cal M} }
\bigg( \sum_{j=1}^{k-1}I(A:B_j|MB_1\ldots B_{j-1})_{\pi}\\
+ \max_{ \Lambda_{k} \in {\cal M} }  I(A:B_k | M B_1 \ldots  B_{k-1})_{\pi}\bigg)  
\end{multline}}

Now
\longshort{\begin{equation} \label{averagevbermu}}{$}
I(A:B_k | M B_1 \ldots  B_{k-1})_{\pi} = \E_{m\sim \mu}  I(A:B_k | B_1 \ldots  B_{k-1})_{\pi_m}.
\longshort{\end{equation}}{$}
Since the $B_1\ldots B_{k-1}$ systems of $\pi_m$ are classical, we can
write the state of $\rho^{AB_k}$ as an average over them, namely
\longshort{\be}{$} \rho^{AB} = \rho^{AB_k} = \sum_i q_i \rho_i^{AB_k},
\longshort{\ee}{$}
where $\{q_i, \rho_i\}$ depend on $\Lambda_{1}, \ldots,
\Lambda_{{k-1}}$ but not on $\cE_m$ and $\Lambda_{k}$.
Then define 
\longshort{\be}{$}
\pi_{i, m}^{AB_k} :=\left(\cE_m^{A\rar \tA} \otimes \Lambda^{B_k} \right)(\rho_{i}^{AB_k}), 
\longshort{\ee}{$}
 so that $\pi_m^{AB_k} = \sum_i q_i \pi_{i, m}^{AB_k}$ and 
\longshort{\begin{equation} \label{measurementaverage}}{$}
I(A:B_k | B_1 \ldots  B_{k-1})_{\pi_m} = \sum_{i} q_{i} I(A : B_k)_{\pi_{i, m}}
\longshort{\end{equation}}{$}
 By Pinsker's inequality \longshort{
\begin{equation} \label{pinsker}
I(A:B_k | B_1 \ldots  B_{k-1} )_{\pi_{m}} \geq  \frac{1}{2} \sum_{i} q_{i} \left \Vert \cE_m^{A\rar \tA} \otimes \Lambda^{B_k} \left( \rho_{i} - \rho_i^{A} \otimes \rho_i^{B_k}    \right) \right \Vert_{1}^2,
\end{equation}
where $\rho_i^{A}$ and $\rho_i^{B_k}$ are the $A$ and $B_k$ reduced states of $\rho_i$.

By convexity of $x^2$ and the trace norm
\begin{eqnarray} \label{eq:boundkterm-first}
I(A:B_k | B_1, \ldots, B_{k-1} )_{\pi_{m}} &\geq&  \frac{1}{2}  \left \Vert \cE_m^{A\rar \tA} \otimes \Lambda^{B_k} \left( \rho^{AB} -  \sum_{i} q_{i} \rho_i^{A} \otimes \rho_i^{B_k} \right)  \right \Vert_{1}^2. \nonumber \\
\end{eqnarray}
Using Eq. (\ref{averagevbermu})}{and the convexities of $x^2$ and the trace norm, we obtain}

\begin{eqnarray} \label{eq:boundkterm}
\shortonly{&&}
\max_{\Lambda_{k} \in {\cal M} }  I(A:B_k | M B_1 \ldots  B_{k-1})_{\pi} 
\shortonly{\\\nn}
&\geq& \frac{1}{2} \max_{\Lambda_{k} \in  {\cal M}  } \E_{m\sim \mu}  \left \Vert \cE_m^{A\rar \tA} \otimes \Lambda^{B_k} \left( \rho^{AB_k} -  \sum_{i} q_{i} \rho_i^{A} \otimes \rho_i^{B_k} \right)  \right \Vert_{1}^2  \\ &\geq&  \frac{1}{2} \min_{\sigma \in \text{SEP}(A:B_k)} \max_{\Lambda_{k} \in  {\cal M}  }  \E_{m\sim \mu} \left \Vert \cE_m^{A\rar \tA} \otimes \Lambda^{B_k} \left( \rho -  \sigma \right)  \right \Vert_{1}^2.\nn
\end{eqnarray}
Note that the second line is independent of  $\Lambda_{1}, \ldots, \Lambda_{{k-1}}$, since only the ensemble $\{ q_i, \rho_i  \}$ depended on them.

From (\ref{eq:expansionCMI}) and (\ref{eq:boundkterm}),
\begin{eqnarray} 
\shortonly{&&}  \max_{ \Lambda_{1}, \ldots, \Lambda_{k} \in  {\cal M}} \longonly{&&}  I(A:B_1\ldots B_k |M)_{\pi} \\
&\geq&  \max_{\Lambda_{1}, \ldots, \Lambda_{{k-1}} \in  {\cal M}}
\sum_{j=1}^{k-1} I(A:B_j|MB_1\ldots B_{j-1})_\pi  \nonumber \\ 
&+&  \frac{1}{2} \min_{\sigma \in \text{SEP}(A:B_k)} \max_{\Lambda_{k} \in  {\cal M}   } \E_{m\sim \mu}  \left \Vert \cE_m^{A\rar \tA} \otimes \Lambda_{k} \left( \rho -  \sigma \right)  \right \Vert_{1}^2.
\nn\end{eqnarray}
Applying the same argument sequentially to all the remaining conditional mutual informations we find
\ba \label{boundtwicemutual}
&\frac{k}{2} \min_{\sigma \in \text{SEP}(A:B)} \max_{\Lambda \in  {\cal M}  } \E_{m\sim \mu}  \left \Vert \cE_m^{A\rar \tA} \otimes \Lambda^B \left( \rho^{AB} -  \sigma^{AB} \right)  \right \Vert_{1}^2 \longshort{\leq}{\nn\\\leq&}
 \max_{ \Lambda_{1}, \ldots,\Lambda_{k}}  I(A:B_1\ldots B_k |M)_{\pi} \leq \ln |\tA|,
\ea
where we used that $\pi^{\tA} = \E_{m\sim \mu} \left(\cE_m^{A\rar \tA}(\rho^A) \right) \in \cD(\tA)$. 
\longshort{Finally by convexity of $x^2$,
\begin{eqnarray}
\left( \min_{\sigma \in \text{SEP}(A:B)} \max_{\Lambda^B \in  {\cal M} } \E_{m\sim \mu}  \left \Vert \cE_m^{A\rar \tA} \otimes \Lambda^B \left( \rho^{AB} -  \sigma^{AB} \right)  \right \Vert_{1} \right)^2 &\leq&  \frac{2 \ln |X|}{k},
\end{eqnarray}
and we are done with the proof of part 1.}{The result follows by the convexity of $x^2$.}

\longshort{\vspace{0.1 cm}
\noindent \textit{Part 2:} The proof of part 2 is mostly the same as that of part 1, and so we only give a brief outline of the changes.  The main change is to omit the maximizations over $\Lambda_1,\ldots,\Lambda_k$, instead using only the fixed measurement $\Lambda$.    We also set $\sigma = \sum_i q_i \rho_i^A \ot \rho_i^{B_k}$ rather than performing a minimization. As a result, the calculations require only time polynomial in the dimensions of the relevant states.

\vspace{0.1 cm}

\noindent \textit{Part 3:} The proof of part 3 is similar to that of part 1, except that we need to make the following replacements.

\begin{tabular}{p{0.35\textwidth}p{0.58\textwidth}}
\toprule
Part 1 & Part 3 \\
\midrule
quantum states $\rho^{AB_1\ldots B_k}$ & non-signaling distributions
$p(x,y_1,\ldots,y_k|a,b_1,\ldots,b_k)$\\[1em]
quantum mutual information & classical mutual information maximized
over choices of measurements $a, b_1,\ldots b_k$ \\[1em]
partial trace & no-signaling condition \\
\bottomrule
\end{tabular}

\vspace{0.1 cm}

For brevity we will use the abbreviations $b^k := (b_1,\ldots,b_k)$,
$b^{k-1} = (b_1,\ldots,b_{k-1})$ and so on.
In more detail, the analogue of \eq{pi-def} is to define the
non-signaling distribution $\pi$ from $\cB^k\rar \cX \times \cY^k$:
\be \pi(x,y^k, a|b^k) = \mu(a) p(x,y^k|a,b^k)\ee
 We can also define
\be \pi(x,y^{k-1},a|b^{k-1}) = \mu(a)p(x,y^{k-1}|a,b^k),\ee
and, thanks to the no-signaling property of $p$, this is well-defined,
since the RHS does not depend on $b_k$.

 Again the chain rule gives us an analogue of \eq{expansionCMI}.
\begin{eqnarray} 
&&\max_{b^k \in \cB^k} I(X:Y^k|A)_{\pi(\cdot | b^k)} \nonumber \\
&=& \max_{b^{k-1}\in \cB^{k-1}} \L(
\sum_{j=1}^{k-1} I(X:Y_j|A Y^{j-1})_{\pi(\cdot | b^{k-1})}
+ \max_{b_k \in \cB} I(X:Y_k | A Y^{k-1})_{\pi(\cdot|b^k)}\R)
\label{eq:NS-chain2}
\end{eqnarray}
Again we focus on the last term of Eq. \eq{NS-chain2}.  Define $i := (a,
b^k, y^{k-1})$, and compute
\ban
\max_{b_k\in \cB} I&(X:Y_k | A Y^{k-1})_{\pi(\cdot|b^k)} 
 = \max_{b_k\in \cB} \E_{a\sim \mu}I(X:Y_k | Y^{k-1})_{p(\cdot|a,b^k)} \\
& = \max_{b_k\in \cB} \E_{a\sim \mu}
\sum_{y^{k-1}} p(y^{k-1}|b^{k-1})
 I(X:Y_k)_{p(\cdot | i)}\\
& = \frac{1}{2}\max_{b_k\in \cB} \E_{a\sim \mu}
\sum_{y^{k-1}} p(y^{k-1}|b^{k-1})
\L(\sum_{\substack{x\in \cX\\y_k\in \cY}}
|p(x,y_k|i) - p(x|i) p(y_k|i)|\R)^2
& \text{Pinsker} \\
& \geq \frac{1}{2}\max_{b_k\in \cB} \E_{a\sim \mu}
\L(\sum_{y^{k-1}} p(y^{k-1}|b^{k-1})
\sum_{\substack{x\in \cX\\y_k\in \cY}}
|p(x,y_k|i) - p(x|i) p(y_k|i)|\R)^2 & \text{convexity of $x\mapsto x^2$}  \\
& \geq \frac{1}{2}\max_{b_k\in \cB} \E_{a\sim \mu}
\L(\sum_{\substack{x\in \cX\\y_k\in \cY}}
\L|p(x,y_k|a,b^k) - 
\sum_{y^{k-1}} p(y^{k-1}|b^{k-1}) p(x|i) p(y_k|i)\R|\R)^2 &\text{convexity of $\|\cdot\|_1$}\\
& \geq 
\frac{1}{2}\min_{q\in \text{LHV}}\max_{b\in \cB} \E_{a\sim \mu}
\|p(X,Y_k | a, b) - q(X, Y_k | a, b)\|_1^2
\ean

As with part 1, we can repeatedly apply this inequality to
\eq{NS-chain2} in order to prove the theorem.
}{The proofs of parts 2 and 3 follow similar lines and can be found in \cite{BH12a}.}
\end{proof}

We conclude this section by comparing our work with previous results on the distance of  $k$-extendable states to separable states.  Our bound, like that of \cite{BCY10b}, requires $k\gtrsim\log |A|$ to achieve constant error in the 1-LOCC norm (ignoring for simplicity the possibility that $|\tilde A| \ll |A|$).  By contrast, previous work~\cite{Ren07,CKMR07,NOP09,DW12} generally required that $k\geq \poly(|B|)$ to achieve constant error in the trace norm.  It is possible to convert from the 1-LOCC norm to the trace norm by giving up a factor of dimension~\cite{MWW09}, but at first it might seem that dependence on $|A|$ is incomparable to dependence on $|B|$.  However, it turns out that our information-theoretic methods can indeed also work in the case where $k\geq\poly(|B|)$, although with polynomially worse scaling than \cite{NOP09}.

\begin{prop}\label{prop:B-dim}
Suppose that $\rho^{AB}$ has a symmetric extension $\tilde \rho^{AB_1\ldots B_k}$.  That is, $\rho^{AB} = \tilde\rho^{AB_i}$ for each $i$ and $\supp\tilde\rho^{B_1\ldots B_k}\subseteq \vee^kB$ (i.e. the symmetric subspace of $B^{\ot k}$.)  Then there exists $\sigma\in\Sep(A:B)$ with 
\be \|\rho-\sigma\|_1 \leq 6 |B|^2\sqrt{\frac{\ln(k)}{k}}.\ee
\end{prop}

It is known~\cite{Ren05} that any extension can be made into a symmetric extension of this form at the cost of squaring $|B|$.  

\begin{proof}
Since $\dim\vee^kB = \binom{|B|+k-1}{k}$ it follows that $I(A:B_1\ldots B_k)_{\tilde \rho} \leq |B|\ln(k)$.  Let $\Lambda$ be an informationally complete measurement on $B$.  This is defined in Lemma 16 of \cite{BH-product} (which itself is a repackaging of \cite{LW12b}), which also proves that 
\be \| \Lambda(X)\|_1 \geq \frac{\|X\|_1}{ \sqrt{18|B|}}\label{eq:info-complete}\ee
 for any matrix $X$.  We will in fact need a variant of this bound:
\be \|(\id_A \ot \Lambda_B)(X)\|_1 \geq \frac{\|X\|_1}{\sqrt{18|B|^3}}\label{eq:info-complete-cb}.\ee
To prove this, 
  define $\Lambda^{-1}$ to be the inverse of $\Lambda$ restricted to the range of $\Lambda$.  Then \eq{info-complete} is equivalent to the statement that $\|\Lambda^{-1}\|_{1\rar 1} \leq \sqrt{18|B|}$ and \eq{info-complete-cb} is equivalent to $\|\id_A \ot \Lambda^{-1}_B\|_{1\rar 1} \leq \sqrt{18|B|^3}$.  Here $\|T\|_{1\rar 1} := \sup_{X\neq 0} \|T(X)\|_1 / \|X\|_1$ for $T$ a linear map on $d\times d$ matrices.  From Lemma 23 of \cite{HaydenW12} and Theorem 11.1 of \cite{Kitaev:02a} we have that for any $d'$,  $\|\id_{d'} \ot T\|_{1\rar 1} \leq \min(d,d') \|T\|_{1\rar 1}$.  Setting $d=|B|$, \eq{info-complete-cb} follows.

The rest of the proof proceeds in a way similar to that of \thmref{bipartite}.
  Let $\pi:=(\id \ot \Lambda^{\ot k})(\tilde\rho)$.  Using first data processing and then the chain rule we find that
\be |B|\ln(k) 
\geq I(A:B_1\ldots B_k)_{\tilde \rho}
\geq I(A:B_1\ldots B_k)_{\pi}
= \sum_{j=1}^k I(A:B_j|B_{<j})_{\pi}.\ee
Thus there exists $j$ for which 
\be I(A:B_j|B_{<j})_{\pi} \leq |B|\frac{\ln(k)}{k}
\label{eq:B-CMI}.\ee

Now consider the state that would result from measuring only $B_1,\ldots,B_{j-1}$ using $\Lambda$.  Call the outcome of this measurement $i$, the post-measurement state $\rho_i^{AB_j}$, and the probability of this outcome $p_i$.  Then $\rho^{AB} = \tilde\rho^{AB_k} = \sum_i p_i \rho_i^{AB_j}$, and $I(A:B_j|B_{<j})_{\pi} = \sum_i p_i I(A:B_j)_{(\id \ot \Lambda)(\rho)}$.  Now using Pinsker's inequality and convexity we find that 
\be \frac 1 2 \| (\id \ot \Lambda)(\rho^{AB} - \sigma^{AB})\|_1^2 \leq 
|B|\frac{\ln(k)}{k},\label{eq:B-almost-there}\ee
with $\sigma := \sum_i p_i \rho_i^A \ot \rho_i^B$.
Applying \eq{info-complete-cb} to \eq{B-almost-there} completes the proof.
\end{proof}

\longshort{\section{Proof of \thmref{multi}}}
{We now turn to the proof of \thmref{multi}}

For a state $\rho^{A_1\ldots B_k}$ we define the multipartite mutual information 
\longshort{\be \label{cmirelent}}{$}
I(A_1:\ldots :A_k) := S(\rho^{A_1\ldots A_k} || \rho^{A_1} \otimes \ldots  \otimes \rho^{A_k}) = S(A_1) + \ldots  + S(A_k) - S(A_1\ldots A_k).
\longshort{\ee}{$}
For a quantum-classical state $\rho^{A_1\ldots A_kR} = \sum_i p_i
\rho_{i}^{A_1\ldots A_k} \otimes \ket{i}\bra{i}^R$ we define the
conditional multipartite mutual information as follows 
\longshort{\be}{$}
I(A_1:\ldots :A_k|R)_{\rho} := \sum_i p_i I(A_1 : \ldots  : A_k)_{\rho_i}.
\longshort{\ee}{$}

\longonly{The multipartite mutual information satisfies the following properties:}
\begin{lem} \label{propertiesmultiMI} \mbox{}
\benum 
\item \text{Multipartite-to-Bipartite \cite{YHHHOS09}:} \shortonly{\vspace{-1ex}}
\ba \label{eq:multitobi}&I(A_1:\ldots :A_k|R) 
\shortonly{\\ \nn &} = I(A_1:A_2|R) 
\longonly{+ I(A_1A_2 : A_3|R)}
 + \ldots  + I(A_1\ldots A_{k-1} : A_k | R).
\ea \shortonly{\vspace{-1ex}}
\item \text{Monotonicity under Local Operations:} Let $\pi^{A_1\ldots A_k} = \Lambda^{A_1} \otimes \id^{A_2\ldots A_k} (\rho^{A_1\ldots A_k})$, then \shortonly{\vspace{-1ex}}
\begin{equation} \label{monotonicitymulti}
I(A_1:\ldots A_k)_{\pi} \leq I(A_1:\ldots :A_k)_{\rho}
\end{equation}\shortonly{\vspace{-1ex}}
\item \text{Pinsker's Inequality:} \shortonly{\vspace{-1ex}}
\begin{equation} \label{eq:pinskerineqmulti}
I(A_1 : \ldots  :A_k)_{\rho} \geq \frac{1}{2} \Vert \rho^{A_1\ldots A_k} - \rho^{A_1} \otimes \ldots  \otimes \rho^{A_k} \Vert_1^2.
\end{equation}\shortonly{\vspace{-2ex}}
\eenum
\end{lem}

\longonly{\begin{repthm}{thm:multi}
\ThmMulti
\end{repthm}}

\begin{proof}\longonly{\mbox{} 

\vspace{0.1 cm}

\noindent \textit{Part 1:}
}
 
Let 
\be \label{pimulti}
\pi^{A_1\ldots A_lR} := (\id^{A_1} \otimes \Lambda_{2} \otimes \ldots  \otimes \Lambda_{l} \otimes {\cal E}^{A_{l+1}\ldots A_k}) (\rho^{A_1\ldots A_k}),
\ee
with $\Lambda_{j} : {\cal D}(A) \rightarrow {\cal D}(X)$ and ${\cal E} : {\cal D}(A^{\otimes k - l}) \rightarrow {\cal D}(R)$ quantum-classical channels. Then from Eq.~(\ref{eq:multitobi})\longonly{ of Lemma \ref{propertiesmultiMI}},
\begin{eqnarray} \label{boundmultiintermsofbipartite}
\shortonly{&&}\min_{ {\cal E}  } \max_{\Lambda_2, \ldots, \Lambda_l} I(A_1 : \ldots  : A_l | R)_{\pi}  
\longshort{&=&}{\nonumber \\ &=&} 
\min_{ {\cal E}  } \max_{\Lambda_2, \ldots, \Lambda_l} \sum_{j=2}^l I(A_1\ldots A_{j-1}:A_j|R)_\pi\nn\\
&\leq&
\min_{ {\cal E}  } \max_{\Lambda_2, \ldots, \Lambda_l} \sum_{j=2}^l I(A_1\ldots A_{j-1}:A_j|R)_{\pi_j}
\label{eq:min-max-sum}\end{eqnarray}
with 
\longshort{\be \label{pildef}}{$}
\pi_{j} :=  (\id^{A_1\ldots A_{j-1}} \otimes \Lambda_j \otimes \id^{A_{j+1}\ldots A_{l}} \otimes {\cal E}^{A_{l+1}\ldots A_k}) (\rho^{A_1\ldots A_k}). 
\longshort{\ee}{$}
The last inequality in Eq. (\ref{boundmultiintermsofbipartite}) follows \longshort{ by the monotonicity of the mutual information under local operations (Eq. (\ref{monotonicitymulti}) of Lemma \ref{propertiesmultiMI})}{from Eq.~(\ref{monotonicitymulti})}. Then
\begin{eqnarray} \label{boundmultiintermsofbipartite2}
\longonly{\shortonly{&&}\min_{ {\cal E}  } \max_{\Lambda_2, \ldots, \Lambda_l} I(A_1 : \ldots  : A_l | R)_{\pi} 
\shortonly{\nn\\} &\leq& 
\min_{ {\cal E}  } \max_{\Lambda_2, \ldots, \Lambda_l} \sum_{j=2}^l I(A_1\ldots A_{j-1}:A_j|R)_{\pi_j}
\nn \\}
\shortonly{\eq{min-max-sum}}
&=& \min_{\cal E} \sum_{j=2}^l \max_{\Lambda_j} I(A_1\ldots A_{j-1}:A_j|R)_{\pi_j} \nn \\
&\leq& \min_{\cal E}  [(l-1) \max_{\Lambda_l}  I(A_1\ldots A_{l-1}:A_l | R)_{\pi_{l}}] , 
\end{eqnarray}
where the last inequality follows from the monotonicity of mutual information \longonly{under tracing out }and the permutation invariance of the state $\rho^{A_1\ldots A_k}$.

We claim that
\begin{equation} \label{boundbipartite}
\min_{ { \cal E }  } \max_{\Lambda_l} I(A_1\ldots A_{l-1} : A_l |
R)_{\pi_l} \leq \frac{(l-1) \ln|A|}{k - l + 1}. 
\end{equation}
Indeed, defining $\nu^{A_1\ldots A_k} := (\id^{A_1\ldots A_{l-1}} \otimes \Lambda_{l} \otimes \ldots  \otimes \Lambda_{k})(\rho^{A_1\ldots A_k})$, for quantum-classical channels $\Lambda_{j}$, we have
\ba
& \max_{\Lambda_l, \ldots, \Lambda_k} I(A_1\ldots A_{l-1} : A_{l} \ldots  A_k)_{\nu} \nonumber  \\ 
=& \max_{\Lambda_l, \ldots, \Lambda_k}
\sum_{j=l}^k I(A_1\ldots A_{l-1}:A_j|A_{j+1}\ldots A_k)_\nu \nn \\
=& \max_{\Lambda_{l+1}, \ldots, \Lambda_k} \Bigg( 
\sum_{j=l+1}^k I(A_1\ldots A_{l-1}:A_j|A_{j+1}\ldots A_k)_\nu 
\shortonly{\nn \\&}
+ \max_{\Lambda_l} I(A_1\ldots A_{l-1} : A_l | A_{l+1} \ldots  A_k)_{\nu} \Bigg) \nonumber \\ 
\geq & \max_{\Lambda_{l+1}, \ldots, \Lambda_k} \Bigg( 
\sum_{j=l+1}^k I(A_1\ldots A_{l-1}:A_j|A_{j+1}\ldots A_k)_\nu 
\shortonly{\nn \\ &}
+ \min_{{\cal E}} \max_{ \Lambda_l} I(A_1\ldots A_{l-1} : A_l | R )_{\pi_l}\Bigg),  
\ea
where the last inequality comes from replacing the specific measurement $\Lambda_{l+1}\ot\cdots\ot\Lambda_k$ with the minimum over all measurements $\cE$ on systems $A_{l+1}\ldots A_k$.
Iterating the argument and exploiting permutation invariance we find
\ba
&\max_{\Lambda_l, \ldots, \Lambda_k} I(A_1\ldots A_{l-1} : A_{l}
\ldots  A_k)_{\nu} 
\longshort{\geq}{\nn\\ \geq  &}
(k - l+1) \min_{{\cal E}} \max_{\Lambda_l} I(A_1\ldots A_{l-1} : A_{l} | R)_{\pi_l},
\ea
and obtain Eq. (\ref{boundbipartite}) from the bound $(l-1) \ln |A| \geq I(A_1\ldots A_{l-1} : A_{l} \ldots  A_k)_{\nu}$. Combining it with Eq. (\ref{boundmultiintermsofbipartite2}) we get
\be \label{boundmultipartitemeasue}
\min_{ {\cal E}  } \max_{\Lambda_2, \ldots, \Lambda_l} I(A_1 : \ldots  : A_l | R)_{\pi} \leq \frac{(l-1)^2 \ln|A|}{k - l + 1}.
\ee

We now show how to combine this bound with a few properties of the measure $I(A_1 : \ldots  : A_l | R)$ to complete the proof. We have
\be 
I(A_1 : \ldots  : A_l | R)_{\pi} = \sum_{i} p_i I(A_1 : \ldots  : A_l)_{\pi_i},
\ee
with $\pi_i := (\id^{A_1} \otimes \Lambda_2 \otimes \ldots  \otimes \Lambda_l) (\rho_i)$, for an ensemble $\{  p_i, \rho_i\}$ such that each $\rho_i \in {\cal D}(A^{\otimes l})$ is permutation-invariant and $\sum_i p_i \rho_i = \rho^{A_1\ldots A_l}$. Then, by Pinsker's inequality (Eq.~(\ref{eq:pinskerineqmulti})) and the convexity of $x^2$: 
\begin{multline*}
\min_{ {\cal E}  } \max_{\Lambda_2,\ldots,\Lambda_l} 
I(A_1 :\cdots  : A_l | R)_{\pi} 
\geq \shortonly{\\}\frac{1}{2} \left \Vert (\id^{A_1} \otimes \Lambda_2
  \otimes \cdots  \otimes \Lambda_l) \left(\rho^{A_1\ldots A_l} -
    \sum_i p_i 
\longshort{\rho_i^{A_1} \otimes \ldots  \otimes \rho_i^{A_l}}{\bigotimes_{j=1}^l \rho_i^{A_j}}
    \right) \right \Vert_{1}^2
\end{multline*}
\longshort{Part 1 of the}{The} theorem follows from Eq. (\ref{boundmultipartitemeasue}).

\shortonly{\end{proof}

The proofs of Corollaries \longonly{\ref{cor:learning},} \ref{cor:separability}, \ref{cor:sos}, \ref{cor:bellqma}, and \ref{cor:hardnessbell} are given in the full version \cite{BH12a}.}

\longonly{
\vspace{0.3 cm}

\noindent \textit{Part 2:}
Let $p(x_1,\ldots,x_k | a_1,\ldots,a_k)$ be a permutation-symmetric
non-signaling distribution and $\mu = \mu_1\times \cdots \times \mu_k$ a product distribution on $A_1 \times A_k$.  We will use the abbreviations $X_{<l} :=
X_1\ldots X_{l-1}$ and $X_{>l}:=X_{l+1}\ldots X_k$.
\begin{subequations}
\label{eq:box-peel-off-one}\ba
 \min_{a_{l+1},\ldots,a_k} \E_{a_1,\ldots, a_l} 
I(X_1:\cdots: X_l |X_{>l})_p 
&=\min_{a_{l+1},\ldots,a_k} \E_{a_1,\ldots, a_l} 
\sum_{j=2}^l I(X_{<j}:X_j|X_{>l})_p \\
&=\min_{a_{l+1},\ldots,a_k} \sum_{j=2}^l\E_{a_1,\ldots, a_l} 
I(X_{<j}:X_j|X_{>l})_p \\
&\leq(l-1)\min_{a_{l+1},\ldots,a_k}\E_{a_1,\ldots, a_l} 
I(X_{<l}:X_l|X_{>l})_p 
\ea\end{subequations}
To derive the last inequality, observe that
$I(X_{<j}:X_j|X_{>l}) = I(X_{<j};X_l|X_{>l}) \leq
I(X_{<l};X_l|X_{>l})$, where the equality is from the symmetry of $p$
and the inequality is from the monotonicity of mutual information
under tracing out systems.

Next,
\begin{subequations}\ba 
(l-1)\ln |X| & \geq 
\min_{a_{l+1},\ldots,a_k} \sum_{j=l}^k \E_{a_1,\ldots,a_l} 
I(X_{<l}: X_j|X_{>j})_p \\
& = 
\min_{a_{l+1},\ldots,a_k} \L(\sum_{j=l+1}^k \E_{a_1,\ldots,a_{l-1}} 
I(X_{<l}: X_j|X_{>j})_p  + 
\E_{a_1,\ldots,a_l} I(X_{<l}: X_l|X_{>l})_p\R)\\
& \geq
\min_{a_{l+1},\ldots,a_k} \sum_{j=l+1}^k \E_{a_1,\ldots,a_{l-1}} 
I(X_{<l}: X_j|X_{>j})_p
+ \min_{a_{l+1},\ldots,a_k}\E_{a_1,\ldots,a_l} I(X_{<l}: X_l|X_{>l})_p
\ea\end{subequations}
Iterating, we find that
\ba
\min_{a_{l+1},\ldots,a_k}\E_{a_1,\ldots,a_l}
 I(X_{<l}:X_l|X_{>l})_p &\leq \frac{(l-1)\ln|X|}{k-l+1}\\
\min_{a_{l+1},\ldots,a_k}\E_{a_1,\ldots,a_l}
 I(X_1:\cdots:X_l|X_{>l})_p &\leq \frac{(l-1)^2\ln|X|}{k-l+1}
&\text{using \eq{box-peel-off-one}}
\label{eq:multi-box-minmax}
\ea
Fix $a_{l+1},\ldots,a_k$ achieving the minimum in Eq. \eq{multi-box-minmax}.
Using the non-signaling property, we can decompose 
\be p(X_{\leq l} | A_{\leq l}) = 
\sum_{x_{>l}} p(x_{>l}|a_{>l}) p(X_{\leq l} | A_{\leq l}, a_{>l},
x_{>l}).\ee
The astute reader will realize that it is now time to deploy Pinsker's
inequality (Eq. \eq{pinskerineqmulti}). Along with Eq. \eq{multi-box-minmax}
and the convexity of $x^2$, this concludes the proof of the theorem.
\end{proof}

\section{Proof of \corref{hardnessbell}}  \label{sec:proof-games}

The first lemma is an adaptation of a similar result of  Kempe, Kobayashi, Matsumoto, Toner, Vidick \cite{KKMTV11}. It shows that by symmetrizing the questions and answers of a subset $S$ of the players one can without loss of generality assume that the players follows a symmetric strategy (in the case of classical, entangled, or non-signaling strategies) 

\begin{lem}\label{symmetrization}
Let $G(N, \pi, V)$ be a non-signaling-prover game such that $\pi(i_1, \ldots, i_N)$ is symmetric in $i_1, \ldots, i_m$ and $V$ is symmetric under simultaneous permutation of registers $1, \ldots, m$ of the questions $q_{i_1,\ldots,i_N}$ and of the answers $a_{i_1,\ldots,i_N}$ for $m \leq N$. Then given any strategy given by a non-signaling strategy that wins with probability $p$, there exists a symmetric strategy with respect to provers $1, \ldots, m$.
\end{lem}

The next lemma gives a hardness of approximation result for approximating the classical value of free games.

\begin{lem} [Aaronson-Impagliazzo-Moshkovitz~\cite{AIM14}] \label{lem:free-games}
3-$\SAT$ with $n$ variables can be reduced to the problem of obtaining a constant error approximation to $\omega_c(G)$ for two-player one-round free games with $2^{O(\sqrt{n})}$-sized output alphabet.
\end{lem}

\begin{repcor}{cor:hardnessbell}
\CorGame{this-text-is-here-to-avoid-multiply-defined-labels}  
\end{repcor}

\begin{proof} \mbox{}

\noindent \textit{Part 1:} Define a game $\overline G$ in which the verifier chooses a pair $(r, q)$ from the distribution $\pi(r, q)$ and sends $r$ to the first prover (let us call it Alice) and the $q$ to one of the other $m$ provers chosen at random  (let us call them Bob 1 to Bob $m$). The verifier does not send a question and does not expect an answer from the remaining Bobs. Then the verifier uses the answers obtained from Alice and the chosen Bob to compute $V(a, b | r, q)$. Applying Lemma \ref{symmetrization} to the case of non-signaling games we can restrict the parties to use non-signaling distributions which are symmetric on the Bobs. Thus

\begin{eqnarray}
\omega_{ns}(\overline{G}) &=& \sup_{p} \sum_{q, r} \pi(r, q)  \sum_{a, b_1, \ldots, b_m} \left( \frac{1}{m} \sum_{k=1}^m V(a, b_k | r, q_k)  \right) p(a, b_1, \ldots, b_m | r, q_1, \ldots, q_m) \nonumber \\
&=&  \sup_{p \in m-\text{Ext}} \sum_{q, r} \pi(r, q) \sum_{a, b} V(a, b | r, q) p(a, b | r, q),
\end{eqnarray}
where the supremum in the last line is taken over all $m$-extendible non-signaling distributions $p$. Then by Theorem \ref{thm:bipartite}
\begin{eqnarray}
&& \sup_{p \in m-\text{Ext}} \sum_{q, r} \pi(r, q) \sum_{a, b} V(a, b | r, q) p(a, b | r, q) \nonumber \\ &\leq& \sup_{s \in \text{LHV}} \sum_{q, r} \pi(r, q) \sum_{a, b} V(a, b | r, q) s(a, b | r, q) + \frac{1}{2} \sqrt{\frac{2 \ln |A|}{m}}.
\end{eqnarray}

In more detail, since the game is free we have that $\pi(r, q) = \pi_1(r) \pi_2(q)$. Then 
\begin{eqnarray}
&& \left |\sum_{q, r} \pi_1(r) \pi_2(q) \sum_{a, b} V(a, b | r, q) \left(p(a, b | r, q) - s(a, b | r, q) \right) \right| \nonumber \\
&\leq& \mathbb{E}_{\pi_1(r)}  \mathbb{E}_{\pi_2(q)}  \left \Vert  p(a, b | r, q) - s(a, b | r, q) \right \Vert_1 \nonumber \\
&\leq&  \mathbb{E}_{\pi_1(r)}   \max_{q \in Q} \left \Vert  p(a, b | r, q) - s(a, b | r, q) \right \Vert_1.
\end{eqnarray}
From theorem \ref{thm:bipartite}
\begin{eqnarray}
\min_{s \in \text{LHV}} \mathbb{E}_{\pi_1(r)}   \max_{q \in Q} \left \Vert  p(a, b | r, q) - s(a, b | r, q) \right \Vert_1 \leq \frac{1}{2} \sqrt{\frac{2 \ln |A|}{m}}.
\end{eqnarray}

\vspace{0.2 cm}

\noindent \textit{Part 2:} Follows from part 1 and the fact that $\omega_{ns}$ can be computed by a linear program \cite{Ito10}.

\vspace{0.2 cm} 

\noindent \textit{Part 3:} Follows from part 1 of this Lemma and part 1 of Lemma \ref{lem:free-games}. 
\end{proof}

\section{Proof of \corref{bellqma}} \label{sec:proof-bellqma}

We begin with a definition of analogues of $\QMA$ with multiple unentangled proofs. 

\begin{definition}
A language $L$ is in $\sf{M}$-$\QMA_n(m, s, c)$ if there exists a polynomial-time implementable two-outcome measurement $\{ M_x, I- M_x \}$ from the class $\sf{M}$ such that
\benum 
\item Completeness: If $x \in L$, there exist $m$ proofs $\sigma_1,\ldots,\sigma_m$, each of $n$ qubits, such that 
\be
\tr\left(  M_x \left(  \sigma_1 \otimes \cdots  \otimes \sigma_m   \right)  \right) \geq c.
\ee
\item Soundness: If $x \notin L$, then for any states $\sigma_1,\ldots,\sigma_m$,
\be
\tr\left(  M_x \left(\sigma_1 \otimes \cdots  \otimes \sigma_m   \right)  \right) \leq s.
\ee
\eenum

If $\sf{M}$ is the class of all polynomial-time implementable two-outcome measurements we denote the complexity class simply by $\QMA_n(m, s, c)$.
\end{definition}

Some examples of classes of measurements that we consider in this paper are:

\benum 
\item $\sf{\textbf{Bell}}$ is composed of measurements $0 \preceq M \preceq I$ of the form
\begin{equation}
M = \sum_{(i_1, \ldots, i_m) \in S} M_{1, i_1} \otimes \ldots  \otimes M_{m, i_m}
\end{equation}
where $\sum_i M_{j, i} = I$ for all $j \in [m]$, and $S$ is a set of $m$-tuples of indices. In words the $m$ subsystems are measured locally giving outcomes $(i_1, \ldots, i_m)$ and the verifier accepts if $(i_1, \ldots, i_m) \in S$. 

\item $\sf{\textbf{LOCC}}_1$ is composed of measurements of the form
\begin{equation}
M = \sum_{i} M_{1, i} \otimes \ldots  \otimes M_{m, i}
\label{eq:LOCC1}\end{equation}
such that $0 \preceq M_{1, i} \leq I$ for all $i$, and $0 \leq \sum_{i} M_{k, i} \leq I$ for every $k \in \{2, \ldots, m\}$.

\item \textbf{$\sf{\textbf{SEP}}$} is composed of measurements $0 \leq M \leq I$ such that
\be
M = \sum_{i} M_{1, i} \otimes \ldots \otimes M_{1, m},
\ee
for positive semi-definite matrices $M_{j, i}$. 
\eenum

See \cite{HM13} for more examples of classes of measurements as well as relations between them.

We will also make use of $\QMA$ with multiple identical proofs:

\begin{definition}
A language $L$ is in $\sf{M}$-$\SymQMA_n(m, s, c)$ if there exists a polynomial-time implementable two-outcome measurement $\{ M_x, I- M_x \}$ from the class $\sf{M}$ such that
\benum 
\item Completeness: If $x \in L$, there exists a proof $\sigma$ of $n$ qubits such that 
\be
\tr\left(  M_x \sigma^{\otimes m} \right) \geq c.
\ee
\item Soundness: If $x \notin L$, then for any state $\sigma$,
\be
\tr\left(  M_x \sigma^{\otimes m} \right) \leq s.
\ee
\eenum

\end{definition}

We now turn to the proof of \corref{bellqma}.

\begin{repcor}{cor:bellqma}
\CorBellQMA
\end{repcor}

\begin{proof} \mbox{} 

\noindent \textit{Part 1:} To simulate a $\BellSymQMA_{n}(m, c, s)$ protocol in $\QMA_{10n^2m^2/\varepsilon^2}(c, s + \varepsilon)$ the verifier receives the proof of $10n^2m^2/\varepsilon^2$ qubits from the prover and consider it as $10nm^2/\varepsilon^2$ blocks of $n$ qubits. Then he symmetrizes all the blocks, traces out all of them except the first $m$ blocks and runs the original $\BellQMA$ protocol on them. It is clear that completeness is not changed. To analyze soundness we use part 1 of Theorem \ref{thm:multi}. 

\vspace{0.2 cm}

\noindent \textit{Part 2:} Follows easily from the previous part. 

\vspace{0.2 cm}

\noindent \textit{Part 3:} To simulate a $\BellQMA_{n}(m, c, s)$ protocol in $\QMA_{10n^2m^4/\varepsilon^2}(c, s + \varepsilon)$ the verifier receives the proof of $10n^2m^4/\varepsilon^2$ qubits from the prover and consider it as $10nm^3/\varepsilon^2$ blocks of $nm$ qubits. Then he symmetrizes all the blocks, traces out all of them except the first $m$ blocks. Then the divides each of these $m$ blocks into $m$ sub-blocks of $n$ qubits. Let us denote the $i$-th sub-block of $j$-th block by $X_{i, j}$. Then the verifier runs the original $\BellQMA$ protocol using the state in subsystems $X_{1, 1}, X_{2, 2}, \ldots, X_{m, m}$ as a proof. It is clear that completeness is not changed. To analyze soundness we use part 1 of Theorem \ref{thm:multi}.  

\vspace{0.2 cm}

\noindent \textit{Part 4:} Follows easily from the previous part. 

\end{proof}

\section{Proofs of Corollary \ref{cor:separability} and \corref{sos}} \label{sec:proof-separability}

\begin{repcor}{cor:separability} 
\CorSeparability{second}
\end{repcor}

\begin{proof}
According to the promise of the weak membership problem, we are given a state $\rho^{A_1, \ldots, A_l} \in {\cal D}(A_1 \otimes \cdots \otimes A_l)$ and wish to determine whether it is separable or $\eps$-far from separable in the LOCC$^{\leftarrow}$ norm.

The idea of the proof is to approximate the set $\Sep(A_1: \cdots:A_l)$ with its $k$-extendible relaxation, for $k$ chosen to give a good approximation guarantee according to \thmref{multi}.  This means introducing systems $X^1,\ldots,X^k$, each of which is composed of $l$ subsystems (i.e. $X^j := X_1^j \cdots X_l^j$ for each $j$, with $X_i^j \cong A_i$ for each $i,j$), and searching for a state $\sigma^{X^1\cdots X^k}$ such that
\benum
\item $\sigma$ is invariant under permutation of the $X^1,\ldots,X^k$ subsystems;
\item and $\sigma^{X_{1}^1 X_2^2 \ldots X_{l}^l} = \rho^{A_1 A_2 \ldots A_l}$.
\eenum
Given a separable $\rho^{A_1 \ldots A_l}$,  such an extension $\sigma^{ X^1 \cdots X^k }$ exists for every $k\geq l$.
We can determine whether such a $\sigma$ exists using semidefinite programming in time polynomial in the overall dimension of $\sigma$.  
If we choose $k = l + 4l^2 \varepsilon^{-2} \sum_j \log|A_j|$, then this will yield the runtime claimed in \eq{sep-runtime-second}.  Moreover, by Theorem \ref{thm:multi} we have that there exists a measure $\nu$ on ${\cal D}(A_1 \otimes \dots \otimes A_l)$ such that
\begin{equation}
\max_{\Gamma_2, \ldots, \Gamma_l \in {\cal M}} \left \Vert \left( \id \otimes \Gamma_2 \otimes \cdots  \otimes \Gamma_l \right) \left( \sigma^{X^1\cdots X^l} - \int \nu(d\omega) \omega^{\otimes l}  \right)      \right \Vert_1 \leq \varepsilon,
\end{equation}
where $\Gamma_2,\ldots,\Gamma_l$ range over all measurements of $X^2,\ldots,X^l$.  Restricting to measurements on the $X_2^2, \ldots,X_l^l$ subsystems (which we denote $\Lambda_2,\ldots,\Lambda_l$) and using the monotonicity of the trace norm under partial trace, we obtain
\begin{equation} \label{eq:boundonewayloccsep}
\min_{\omega \in \text{Sep}(A_1: \cdots:A_l)} \max_{\Lambda_2, \ldots, \Lambda_l \in {\cal M}} \left \Vert \left( \id \otimes \Lambda_2 \otimes \cdots  \otimes \Lambda_l \right) \left( \sigma^{X_1^1\cdots X_l^l} - \omega \right)      \right \Vert_1 \leq \varepsilon.
\end{equation}
Of course, $\sigma^{X_1^1\cdots X_l^l}$ in \eq{boundonewayloccsep} is equal to $\rho^{A_1\ldots A_l}$, and so the existence of the symmetric extension $\sigma$ implies that $\rho$ is no more than $\eps$-far from separable in the one-way LOCC norm.  Conversely, if $\rho$ is more than $\eps$-far from separable in the LOCC$^{\leftarrow}$ norm, then such a $\sigma$ will not exist, and thus our algorithm will be able to distinguish this case from the one where $\rho$ is separable.

The bound for $\Vert * \Vert_{2(l)}$ follows from the reasoning above and the following bound (given by Theorem 5 of \cite{LW12b}): 
\be
\Vert X \Vert_{\text{LOCC}^{\leftarrow}}  \geq 18^{- l/2} \Vert X \Vert_{2(l)}.
\ee
\end{proof}

\begin{proof}[Proof of \corref{sos}]
Directly applying part 1 of \thmref{multi} to this problem would yield an approximation of $\max_\sigma \tr M\sigma^{\ot l}$.  To instead relate this to $h_{\Sep(A^{\ot l})}(M)$ we introduce systems $A_i^j$ for $i\in[k]$ and $j\in [l]$, for $k$ to be determined later.  Now consider $M$ to be an operator on systems $A_1^1 A_2^2 \cdots A_l^l$, and define $\tilde M$ to be the operator acting on $\bigotimes_{i,j} A_i^j$ with $M$ acting on $A_1^1 \cdots A_l^l$ and $I$ acting on the other positions.

Define the $n^l$-dimensional block systems $B_i := A_i^1\cdots A_i^l$. Let $\rho$ be a permutation-invariant state on $\cD(B_1 \ot \cdots \ot B_k)$.
According to part 1 of \thmref{multi}, there exists a measure $\nu$ over density matrices such that, if $\omega = \int \nu(d\sigma) \sigma^{\ot l}$, then
\be \sqrt{\frac{2l^2\cdot l\ln(n)}{k-l}} \geq
|\tr(\tilde M(\rho^{B_1\ldots B_l}-\omega))|
= |\tr (M(\rho^{A_1^1 \ldots A_l^l} - \omega^{A_1^1 \ldots A_l^l}))|.
\label{eq:multiq-approx}\ee
The maximum of $\tr \tilde M\omega$ over all $\omega$ is given by $h_{\Sep(A^{\ot l})}(M)$.
According to \eq{multiq-approx}, and the fact that one possible solution for $\rho$ is given by $(\bigotimes_{i=1}^l \proj{z^{(i)}})^{\ot m}$, we have
\be h_{\Sep(A^{\ot l})}(M) \leq \max_\rho \tr \tilde M\rho \leq
h_{\Sep(A^{\ot l})}(M) + 
\sqrt{\frac{2l^3\ln(n)}{k-l}}\label{eq:multiq-bounds}.\ee

Finally, optimizing over $\rho$ is a strictly weaker relaxation of $h_{\Sep}$ than is provided by the SOS hierarchy.  That hierarchy would also have constraints such as the PPT condition.  See Lemma 9.10 of \cite{BHKSZ12} for a similar comparison.  Thus, the SOS hierarchy achieves bounds at least as tight as those of \eq{multiq-bounds}.  We conclude by taking $k = O(l^3\ln(n)/\varepsilon^2)$.  Since each $B_i$ is a collection of $l$ basic variables, this corresponds to level $O(l^4 \ln(n)/\ep^2)$ of the SOS hierarchy.  The runtime bound follows from the fact that the total dimension of the resulting SDP is $n^{kl}$.
\end{proof}

\section{Open Problems}

It would be desirable to strengthen several of the results in this work:

\begin{enumerate}

\item Conjecture~\ref{con:games} is a proposed improvement of \thmref{bipartite} that would imply that $O(\log(k))$-extendable states cannot be distinguished from separable states by Bell measurements with $k$ outcomes per party.  As we discuss in Section \ref{sec:games} this would have a very interesting application to the complexity of non-local games.

\item We would also like to improve \thmref{bipartite} to apply to separable measurements\footnote{Technically, we refer here to approximating $h_{\Sep}(M)$ for  ``SEP-YES'' measurements, meaning that $M$ is of the form $\sum_i A_i \ot B_i$ for p.s.d. $A_i,B_i$, but without any such requirement for $I-M$.} instead of merely 1-LOCC measurements.  If this were true, it would imply by the results of \cite{HM13}, that $\QMA_n(m,c,s) \subseteq \QMA_{O(mn^2/\eps)}(1,c,s+\eps)$.  It would also yield quasipolynomial-time classical algorithms for separability testing and a large number of tensor optimization problems described in \cite{HM13}.

\item One of the few barriers to improving de Finetti theorems is the example of the maximally mixed state on the antisymmetric subspace of $\bbC^d \ot \bbC^d$~\cite{CKMR07}.  This so-called ``universal counterexample'' state is $d$-extendable, and yet is far from separable.  However, this distinguishability cannot be achieved by a measurement whose ``not separable'' outcome is itself a separable measurement operator; aka a ``SEP-YES'' measurement.  As mentioned in the previous open problem, proving a more efficient de Finetti theorem against such measurements would improve the algorithm for approximating $h_{\text{Sep}}(M)$ for general measurements $M$.
Additionally, the antisymmetric state is not PPT, and such examples of
highly-extendable far-from-separable states are not known to occur
when we add the PPT constraint, as proposed by \cite{DohertyPS04}.
Intriguingly, the ``worst'' known example (i.e. most extendable while
being far from separable) of a PPT state is only $O(\log
d)$-extendable~\cite{BCY10} \footnote{In more detail this follows by considering a variant of the example of \cite{BCY10} (page 6) in which the EPR pair is replaced by a constant-dimensional PPT entangled state. Note that by \cite{BW10} for every $\varepsilon > 0$ there is a PPT state with trace distance $2 - \varepsilon$ from separable states, thus for every $\varepsilon > 0$ one can get a PPT $O(\log(d))$-extendible $d \times d$ state which is $(2 - \varepsilon)$-away from any separable state. The same holds true if we use the 1-LOCC version of the trace norm.}.  It would be of great interest either to
prove a better bound on the combination of PPT and $k$-extendable constraints (see \cite{NOP09} or Section 9.3.2 of \cite{BHKSZ12} for some progress), or to find better
counterexample states.

\item It would also be interesting to use our information-theoretic
techniques to examine the various extensions of the de Finetti
theorem.  For example, is there a version of the post-selection
technique~\cite{CKR09} where the dimension dependence is replaced by
a dependence on the number of measurement outcomes?  One difficulty
here (highlighted by taking the local dimension to be infinite) is in
choosing the right test state upon which the channels 
should act.  Another question is whether our techniques can improve
the exponential de Finetti theorem~\cite{Ren05}.  Unfortunately, this
theorem is known not to have a classical analogue (due to unpublished
work of Christandl and Toner), while our proofs use entropic properties of classical, or classical-quantum, states.

\end{enumerate} 

\section*{Acknowledgments}
We are grateful to Kevin Milner, Thomas Vidick and Mark Wilde for many helpful comments on an early version of the paper, to Ashley Montanaro for explaining to us the remark at the end of \secref{BellQMA}, to Scott Aaronson for telling us about \cite{AIM14} in 2010, to Graeme Smith and Ke Li for catching a bug in an earlier version of \corref{sos} and especially to Boaz Barak, Jon Kelner and David Steurer for sharing with us an early version of \cite{BKS12}. We also benefited from interesting discussions with Matthias Christandl and Stephanie Wehner.  Much of this work was done while FGSLB was working at the Institute for Theoretical Physics in ETH Z\"urich and AWH was working in the Department of Computer Science at the University of Washington.
FGSLB acknowledges support from EPSRC through an Early Career Fellowship, the Polish Ministry of Science and Higher Education Grant No. IdP2011 000361, the Swiss National Science Foundation, via the National Centre of Competence in Research QSIT, the German Science Foundation (grant CH 843/2-1), the Swiss National Science Foundation (grants PP00P2$\textunderscore$128455, 20CH21$\textunderscore$138799 (CHIST-ERA project CQC)), the Swiss National Center of Competence in Research "Quantum Science and Technology (QSIT)", and the Swiss State Secretariat for Education and Research supporting COST action MP1006. AWH was funded by NSF grants 0916400, 0829937, 0803478, 1111382, DARPA QuEST contract FA9550-09-1-0044 and ARO contract W911NF-12-1-0486.
}

\end{document}